\documentclass[twocolumn]{autart}

\usepackage{amsmath,amssymb,bm}
\usepackage{graphicx}
\usepackage{cite,natbib}
\usepackage{float}
\usepackage{dsfont}
\usepackage{enumitem}

\usepackage{color}

\newcommand{\blue}[1]{{\color{black}#1}}

\newcommand{\mat}[1]{\left[\; \begin{matrix} #1 \end{matrix} \:\right]}
\newcommand{\simode}[1]{\left\{\;
\begin{aligned} #1 \end{aligned} \right.}

\DeclareMathOperator{\sfker}{\mathsf{ker}}
\DeclareMathOperator{\sfim}{\mathsf{im}}

\DeclareMathOperator{\sfdiag}{\mathsf{diag}}

\DeclareMathOperator{\sfcol}{\mathsf{col}}

\DeclareMathOperator{\sfdet}{\mathsf{det}}

\usepackage{theorem}
\theorembodyfont{\rmfamily}

\newtheorem{theorem}{Theorem}
\newtheorem{lemma}{Lemma}

\newtheorem{proof}{Proof}

\newtheorem{corollary}{Corollary}

\begin{document}
\setlength{\parskip}{4.5pt}
\setlength{\abovedisplayskip}{6pt}
\setlength{\belowdisplayskip}{6pt}
\setlength{\abovedisplayshortskip}{5pt}
\setlength{\belowdisplayshortskip}{5pt}

\begin{frontmatter}

\title{Hierarchical Distributed Architecture for the Least Allan Variance Atomic Timing 
}

\author[TokyoTech]{Jiayu Chen}\ead{jiayuc@lim.sc.e.titech.ac.jp},    
\author[Gunma]{Takahiro Kawaguchi}\ead{kawaguchi@gunma-u.ac.jp},               
\author[NICT]{Yuichiro Yano}\ead{y-yano@nict.go.jp},  
\author[NICT]{Yuko Hanado}\ead{yuko@nict.go.jp},
\author[TokyoTech]{Takayuki Ishizaki}\ead{ishizaki@sc.e.titech.ac.jp}

\address[TokyoTech]{Institute of Science Tokyo, Meguro, Tokyo, 152-8552, Japan}  
\address[Gunma]{Gunma University, Kiryu, Gunma, 371-8510, Japan}             
\address[NICT]{National Institute of Information and Communications Technology, Koganei, Tokyo, 184-0015, Japan}        



\begin{abstract}
In this paper, we propose a hierarchical distributed timing architecture based on an ensemble of miniature atomic clocks. 
The goal is to ensure synchronized and accurate timing in a normal operating mode where Global Navigation Satellite System (GNSS) signals are available, as well as in an emergency operating mode during GNSS failures. 
At the lower level, the miniature atomic clocks employ a distributed control strategy that uses only local information to ensure synchronization in both modes. 
The resulting synchronized time or generated time scale has the best frequency stability, as measured by the Allan variance, over the short control period. 
In the upper layer, a supervisor controls the long-term behavior of the generated time scale. 
In the normal operating mode, the supervisor periodically anchors the generated time scale to the standard time based on GNSS signals, while in the emergency operating mode, it applies optimal floating control to reduce the divergence rate of the generated time scale, which is not observable from the measurable time difference between the miniature atomic clocks.
This floating control aims to explicitly control the generated time scale to have the least Allan variance over the long control period.
\blue{
Finally, numerical examples are provided to show that the proposed architecture achieves an Allan variance on the order of $10^{-23}$ over averaging times ranging from one second to several days.
This demonstrates its effectiveness and feasibility for high-precision, GNSS-resilient atomic timing.
}
\end{abstract}
\end{frontmatter}

\section{Introduction}
\subsection{Background}\label{sec:back}
Accurate time synchronization plays a critical role in a wide range of modern distributed systems, including telecommunications, power grids, and financial transactions \cite{bregni2002synchronization,zhang2013time}. 
Currently, most systems rely on mainstream time protocols, such as the Network Time Protocol (NTP) \cite{mills1991internet} and the Precision Time Protocol (PTP) \cite{serrano2013white}, which typically use a tiered server-client model.

In standard NTP operation, a client obtains time information from an NTP server and synchronizes its time to the server, which becomes a server for its own clients. NTP supports up to 15 such hierarchy layers, called {\it stratum}, but synchronization accuracy decreases with each additional stratum \cite{banerjee2023introduction}. 
At the very top, a server in stratum $1$ or a grand master clock synchronizes its system time with highly accurate devices, typically receivers of Global Navigation Satellite System (GNSS) signals \cite{montenbruck2020comparing}.
GNSS is a general term used to describe any constellation of satellites that provide positioning, navigation, and timing services on a global or regional basis. 
Each GNSS satellite is equipped with multiple on-board atomic clocks \cite{jaduszliwer2021past} and continuously broadcasts a GNSS time, which closely approximates the national standard time. 
This allows receivers to access accurate national standard time information within a few hundred nanoseconds \cite{coleman2020autonomous}.

However, this reliance on GNSS has raised significant concerns about system resilience and robustness in recent years \cite{zidan2020gnss,kunzi2023precise}.
GNSS signals are highly sensitive to natural phenomena such as ionospheric scintillation and solar flares. 
In particular, intense solar flares lasting several hours can degrade timing accuracy by two orders of magnitude \cite{pi2017effects}. The increasing frequency of malicious attacks, such as jamming and spoofing \cite{liu2020secure,radovs2024recent}, also poses a serious threat to the reliability of GNSS-dependent timing. 
As emerging applications, such as 6G and beyond communications, industrial IoT, and cooperative autonomous vehicles \cite{hasan2018time,yiugitler2020overview}, demand ever-increasing timing accuracy and security, it is imperative to develop a resilient timing architecture.
It should be noted that in the current timing service, if the grand master clock can no longer receive GNSS signals, the time of all clients will be lost.

To support the emerging demand, the timing system should be able to operate effectively in two basic operations, i.e., normal operation when GNSS signals are available and emergency operation in the event of  GNSS failures.
Specifically, the general \textbf{Requirements} are listed as follows. 
\begin{enumerate}[label=\Roman*.]
  \item \textbf{Consistent Synchronization}: The clocks should remain synchronized in both operating modes.
  \item \textbf{Accuracy to National Standard Time}: The generated time scale (GTS), i.e., the synchronized time shared by all clocks, should have a small deviation from the national standard time in both operating modes.
\end{enumerate}

A promising approach to meet these requirements is the use of miniature atomic clocks (MACs) \citep{kitching2018chip}. 
Advances in chip-scale atomic clock technology \cite{cash2018microsemi} have shown that they can provide excellent stability and reduced drift, which is the most important property for timing devices. 
In particular, the Allan variance (AVAR) \citep{allan1966statistics} is used as a standard measure to evaluate frequency stability over time in oscillators, which calculates the variance of the differences between successive averages of frequency measurements over specified sampling intervals.  
This measure differs from directly measuring the variance of the frequency or its derivative, because it focuses on evaluating the changes in frequency over these period averages rather than examining the instantaneous frequency fluctuations. 
In other words, clocks with a small AVAR over long averaging periods are able to maintain long-term stable timing.
A typical free-running (uncontrolled) MAC exhibits an AVAR of $10^{-12}$ over averaging times of $1000$ seconds or more, whereas conventional quartz clocks exhibit about $10^{-6}$ \cite{gill2005optical,bandi2024comprehensive}. Such long-term stability allows the MACs to be highly effective even in emergency operation.

\subsection{Proposed System Architecture}
\begin{figure*}[!t]
\centering
\includegraphics[scale=0.65]{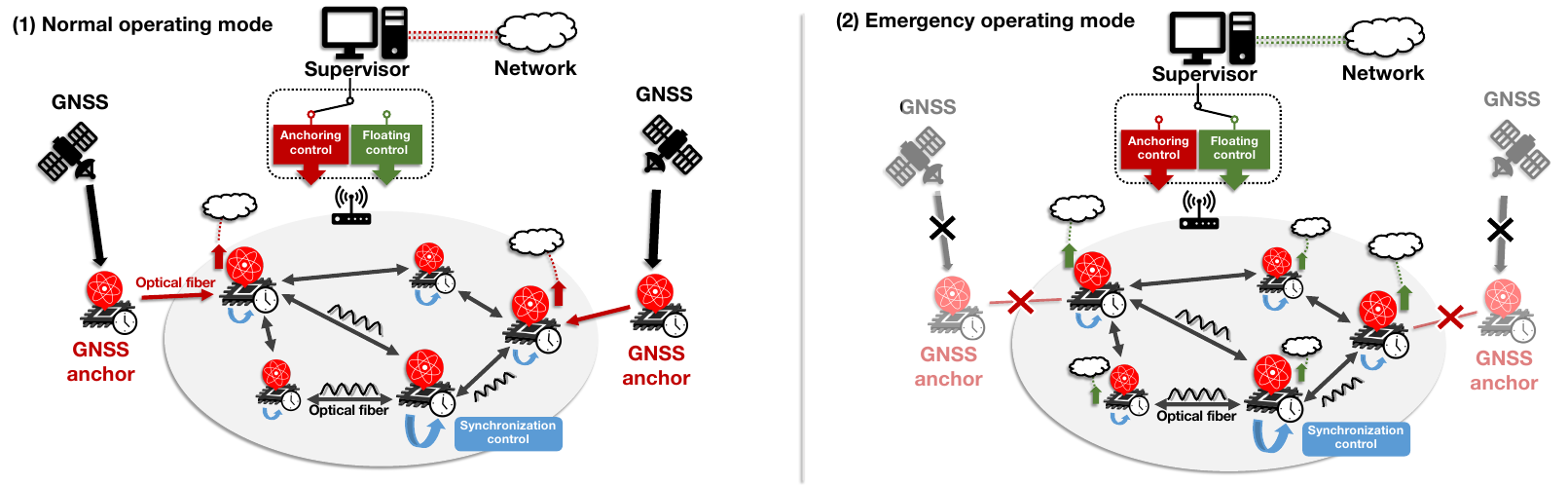}
\caption{Diagram of the hierarchical atomic timing architecture in normal and emergency operation. 
At the lower level, the MACs provide GNSS-independent distributed synchronization, while at the upper level, the supervisor controls the GTS by switching between anchoring and floating control in response to evolving situations.}
\label{fig:archi3}
\end{figure*}

In this paper, we develop a novel distributed atomic timing system based on MACs that are less dependent on GNSS. 
The system adopts a hierarchical architecture as shown in Fig.~\ref{fig:archi3}. 
The lower layer consists of an ensemble of MACs and several GNSS anchors, each also equipped with a MAC and capable of receiving a GNSS signal with a tiny delay. 
These devices are assumed to be deployed in geographically dispersed locations, e.g.,  20-100 km apart in a real use case,
and connected by a fiber optic network \cite{winzer2018fiber}. 
A supervisor locating at the upper layer manages the synchronized
time of all MACs or the GTS by aggregating information from the MACs to regulate their long-term performance less frequently.
The authors' research group is developing a prototype that covers an area of $400~{\rm km^2}$ with 10 MACs, integrating both software and hardware.

Each MAC has two types of control, a frequent short-period control that continuously steers the MAC towards synchronization, and an intermittent long-period control that gradually regulates the GTS.
The former is generated by the individual MACs in a distributed manner using only local information. 
The optical fibers allow the clock reading signals to be exchanged with negligible delay. 
Note that due to a physical limitation of MACs, only the difference between two signals can be measured, not the absolute value of the phase or frequency. 
Each MAC measures the time differences between its adjacent neighbors and itself to 
\blue{
estimate their phase and frequency difference locally using a Kalman filter \cite{anderson2005optimal}
}, and uses a consensus-type controller \cite{olfati2004consensus,chen2019bit} to stabilize these offsets. 
In particular, the synchronization control gain is designed to take into account the special characteristics of the MACs so that the GTS has the least AVAR over the short control period.
The latter type of control is received from the supervisor, which varies depending on the following operating modes.
\begin{itemize}
  \item In the normal operating mode, the GTS is anchored based on the GNSS anchors. 
  The reliability of the system is improved by using multiple anchors because the system can maintain timing even if some of the GNSS anchors are lost.
  Each MAC adjacent to the anchor measures and estimates its deviation from the anchor, and reports the result to the supervisor. 
  Based on the average of all these estimates, the supervisor periodically broadcasts an anchoring control signal to all the MACs to correct their common deviation from the standard time.
  \item In the emergency operating mode, the GTS is controlled according to the concept of ``optimal long-term floating".
  Although neither the GTS nor its optimum is observable over the long control period, it can be shown that their difference can be controlled indirectly based on the measurable inter-clock differences.
  Using this fact, the supervisor collects a subset of the inter-clock estimates from the MACs, and broadcasts an optimal floating control signal to eliminate this difference without the reference to GNSS anchors.
\end{itemize}
It is shown that the GNSS-independent synchronization among the MACs is ensured regardless of the operating modes, while the robustness to GNSS signal uncertainties is greatly enhanced by switching the control modes.
Consequently, the resultant system perfectly satisfies Requirements I and II.
To the authors' knowledge, this is the first attempt to study a hierarchical distributed timing architecture based on the use of MACs.

\subsection{Review of Related Studies}
\subsubsection{Clock Synchronization for Sensor Networks}
While there are similarities to common clock synchronization problems, the clock model and system requirements are quite different.
The problem of distributed clock synchronization has been actively studied, especially in the context of wireless sensor networks \cite{kadowaki2014event,he2017accurate,wang2022consensus}. 
\blue{
They often model clocks with constant bias and drift parameters assuming that the parameters are fully controllable and observable.
They develop distributed consensus algorithms that force the clocks to eventually converge on a common time coordinate.  
However, the resulting time accuracy relative to standard time is not part of the problem formulation.
The focus is usually on dealing with the unavoidable, yet uncertain transmission time delays.

In contrast, the synchronization problem for the MACs in this paper differs significantly from the common ones.
The transmission delay through optical fibers is generally negligible, so it is not a primary concern in designing the control strategy.
Instead, the goal is to maintain a physically meaningful time scale with high accuracy and AVAR performance among the MACs. 
}
Unlike conventional quartz clock models, the difference between the MACs results from the accumulation of their inherent process noise rather than from local parameter mismatches. 
Therefore, a finer control that takes into account the stochastic nature of the MACs is required to maximize the long-term frequency stability of the resulting GTS, especially in emergency operation. 
\blue{
This approach is not included in standard consensus methods.
}

\subsubsection{Standard Time Scale Generation}
The design of a timing architecture that maximizes frequency stability in this paper is closely related to the problem of generating national standard time. 
A centralized scheme of time scale generation based on an ensemble of atomic clocks has been studied in recent decades \cite{galleani2003use}.
By integrating measurements from multiple clocks, an ensemble time, called the implicit ensemble mean \cite{brown1991theory}, is generated that has stability equal to or better than the best individual clock in the ensemble.

Typically, a standard Kalman filter is applied directly to estimate the deviations of the clock readings from the ideal time. 
However, as pointed out by \cite{galleani2010time}, these methods suffer from structural observability limitations and covariance divergence.
\blue{
To address this issue, a basis transformation approach is proposed by the authors’ research group \cite{ishizaki2026explicit} in the centralized setting, where the unobservable GTS is explicitly isolated to enable determinate time-scale regulation. 
In contrast, the focus of this paper is on time-scale generation in distributed MAC ensembles, where only local inter-clock measurements are available. 
We propose a hierarchical distributed timing architecture that enables local synchronization and global GTS adjustment within a unified framework.
This aspect has not been explicitly addressed in existing studies.
}

The rest of the paper is organized as follows. 
In Section \ref{sec:Prob}, we formulate the fundamental problems for atomic timing. 
Sections \ref{sec:Architecture1}  and \ref{sec:Architecture2} propose the hierarchical distributed architecture. 
In particular, the synchronization, anchoring, and floating control strategies are presented in Section \ref{sec:Sync}, \ref{sec:Anchoring}, and \ref{sec:Floating}, respectively, along with illustrative numerical examples.
Section \ref{sec:Conclusion} concludes the paper.

\section{Problem Statement}
\label{sec:Prob}
\subsection{Notations} 
In this paper, we denote
the set of real numbers by $\mathbb R$, 
the set of integers by $\mathbb Z$,
the stationary Gaussian processes with zero-mean and bounded variance by $\mathds G$,  
the $n$-dimensional identity matrix by $I_n$, 
the $n$-dimensional all-ones vector by $\mathds{1}_n$,  
the $i$th column of $I_n$ by \blue{$\bm{\e}_{n|i}$},
the cardinality of a set $\mathcal{N}$ by $|\mathcal{N}|$, 
the set of eigenvalues of a square matrix $A$ by $\bm{\Lambda}(A)$, 
the image of a matrix $A$ by $\sfim A$, 
the kernel of a matrix $A$ by $\sfker A$, 
the expectation of a random variable $X$ by $\mathbb{E}[X]$, 
and 
the Kronecker product of two matrices $A$ and $B$ by $A \otimes B$.
For scalars or matrices $x_i$ for $i \in \mathcal N$, we denote the diagonal or block diagonal matrix of $x_i$ as $\sfdiag (x_i)_{i \in \mathcal N}$, 
\blue{the elementwise stacking of $x_i=[x_{i1},x_{i2},\ldots,x_{in}]^{\sf T} \in \mathbb R^n$ by $ 
\sfcol (x_i)_{i \in \mathcal N} := [x_{11},x_{21},\ldots,x_{|\mathcal N|1},x_{12},x_{22},\ldots,x_{|\mathcal N|n}]^{\sf T}$. }
The subscript ${i \in \mathcal N}$ may be omitted for notational simplicity without causing confusion.

\subsection{System Modeling}
\label{sec:Model}
Consider an ensemble of MACs, whose label set is denoted by $\mathcal{N}:=\{1,\ldots,N\}$. 
We denote the variables by the subscript $i$ to indicate their correspondence to the $i$th MAC, unless otherwise noted. 
Experimental observations \cite{galleani2008tutorial} show that the free-running time series of their clock reading deviations relative to an ideal time reference, denoted by $h_i(t)$, can be modeled as
\blue{
\begin{equation*}
h_i(t) = c_{i1} + c_{i2} t + \int_0^t v_{i1}(t_1) {\text d}t_1 + \int_0^t \int_0^{t_1} v_{i2}(t_2) {\text d} t_2 {\text d} t_1,
\end{equation*}
where $c_{i1}$ and $c_{i2}$ are scalar constants, $v_{i1}(t) $  and  $v_{i2}(t)$ are white frequency noise and random walk frequency noise, whose standard deviations are $\sigma_{i1}$ and $\sigma_{i2}$, respectively. 
This integral equation model can be represented by a second-order stochastic system as
\begin{equation}\label{eq:cmodel}
\simode{
	\dot{x}_{i1}(t) &= x_{i2}(t) + v_{i1}(t), \\
	\dot{x}_{i2}(t) &= v_{i2}(t), \\
	h_i(t) &= x_{i1}(t),
}
\end{equation}
where $x_{i1}(t) \in \mathbb{R}$ represents the clock 
reading deviation and $x_{i2}(t) \in \mathbb{R}$ represents the fractional frequency deviation, with initial values $x_{i1}(0) = c_{i1},~x_{i2}(0) = c_{i2}$.

Consider a time sequence $\{t_k\}_{k \in \mathbb{Z}}$ from the ideal time coordinate with a constant period $\tau$, i.e., $t_k = k \tau$. 
Then a control-oriented discrete-time model that is equivalent
to the continuous-time model in \eqref{eq:cmodel} over the discrete time
sequence is derived as
\begin{equation}\label{eq:dmodel}
\simode{
x_i[k+1] &= A x_i[k] + B u_i[k] + v_i[k],\\
h_i[k] &= C x_i[k],
}
\end{equation}
where $x_i[k] := [x_{i1}(t_k),x_{i2}(t_k)]^{\sf T}$, 
$u_i[k] \in \mathbb{R}$ denotes the steering control input that is used to tune the oscillator inside the MAC \cite{koppang2016state,marlow2021review}, and
}
\[
A = \left[\; \begin{matrix} 
1  & ~\tau \\
0 & ~1
\end{matrix} \:\right],\quad
B = \left[\; \begin{matrix} 
\tau \\
 1
\end{matrix} \:\right],\quad
C = [\; \begin{matrix} 
1 & ~0
\end{matrix} \: ].
\]
The covariance matrix of the white Gaussian noise $\{v_i[k]\}_{k\in \mathbb{Z}}$ is given as
\begin{equation}\label{eq:Q_def}
Q_i(\tau) :=
\mat{
\tau \sigma_{i1}^2 + \frac{\tau^3}{3} \sigma_{i2}^2   & ~\frac{\tau^2}{2}\sigma_{i2}^2  \\
\frac{\tau^2}{2} \sigma_{i2}^2 & ~\tau \sigma_{i2}^2 
},
\end{equation}
which is dependent on the sample period.
Each MAC is connected to several neighbors via bidirectional communication channels. 
Note that the deviation $h_i[k]$ is defined relative to an ideal time reference, so it cannot be measured directly.
Instead, the phase difference between two MACs is measurable. 
Specifically, the measurement made by MAC $i$ relative to MAC $j$ is described as 
\begin{equation}\label{eq:measure_ij}
y_{ij}[k] := h_j[k] - h_i[k] + w_{ij}[k], 
\end{equation}
where $\{w_{ij}[k]\}_{k\in \mathbb{Z}}$ is the white Gaussian measurement noise with a variance $R$.

In addition to the network of MACs, we consider a set of anchors labeled by $\mathcal{M}:=\{1,\ldots,M\}$, which contain information of a standard time. 
Each anchor is equipped with an MAC with dynamics
\begin{equation}\label{eq:dmodel_anc}
X_l[k+1] = A X_l[k] + B U_l[k] + \mu_l[k],
\end{equation}
where $X_l[k]$ denotes the internal state, $U_l[k]$ denotes the steering control input, and $\mu_l[k]$ is the process noise.
The anchor continuously receives the standard time signal from GNSS. 

However, limited by both the broadcast ephemeris accuracy and pseudorange measurement errors, the received signal is corrupted, including a constant offset and large measurement noise. 
With such imprecise information, the anchor gradually adjusts its internal state to the standard time.
Steering methods such as \cite{dey2024clock} can be applied to ensure a relatively stable output provided by an anchor. 
Considering this, we assume that the state of the $l$th anchor satisfies 
\begin{equation}\label{eq:Exanc}
\mathbb{E}[X_l[k]] = [\; \begin{matrix} 
\Theta_l^* & ~0
\end{matrix} \: ]^{\sf T}
\end{equation} 
and has a variance $Q^a_l(\tau)$, which takes the form of \eqref{eq:Q_def} parameterized by $\sigma_{l1}^a$ and $\sigma_{l2}^a$.

Each anchor is connected to several MACs in the ensemble. 
Denote the label set of MACs neighboring the $l$th anchor by $\mathcal{W}_l$.
When the connections are active, such a MAC can measure their phase difference as
\begin{equation}\label{eq:measure_anchor}
Y_{li}[k] := C (X_l[k] - x_i[k]) + \omega_{li}[k],\quad i \in \mathcal{W}_l,
\end{equation}
which contains information about the deviation from the standard time.

\subsection{The Allan Variance}
The Allan variance (AVAR) \cite{ishizaki2023higher} is one of the most common measures of frequency stability for clocks, oscillators and amplifiers. 
\blue
{
Suppose that $u[k]$ is zero for a single MAC model as described in \eqref{eq:dmodel}.
}
The AVAR is defined as
\begin{equation}\label{eq:allan_def}
\sigma_{\sf A}^2 \bigl( m \tau \bigr) := \mathbb{E} \left[ \frac{ \left( \triangle_{m}^{2} h_i[k] \right)^2 }{2 (m\tau)^2} \right],
\end{equation}
where the second-order difference operator is denoted by
\[\triangle_{m}^{2}h_i[k]:=(h_i[k+2m] - h_i[k+m]) - (h_i[k+m] - h_i[k])
\]
with step $m \geq 1$.
It is intended to assess stability due to noise processes. 
\blue{
In fact, the Allan variance of the free-running clock model is a function only of $\tau$. Its analytical form can be found \cite{ishizaki2023higher} as 
\begin{equation}\label{eq:allan_ana}
\sigma_{\sf A}^2\bigl( m \tau \bigr) = \tfrac{1}{m\tau} \sigma_{i1}^2 + \tfrac{m\tau}{3} \sigma_{i2}^2.
\end{equation}
On the other hand, when $u[k]$ is applied, the Allan variance is no longer a function of $\tau$ alone, and in general admits no analytical expression. 
}
In such a case, AVAR is often estimated using sampled data.
In particular, for a clock reading sequence 
$\{h_i[0],h_i[1],\ldots,h_i[M]\}$ measured over the sampling period $\tau$, a statistical estimate of the AVAR can be calculated as
\begin{equation}\label{eq:allan_est}
\hat{\sigma}_{\sf A}^2 \bigl(\{h_i[k]\}, m \tau \bigr) := \frac{1}{M - 2m}\!\! \sum_{k=0}^{M - 2m -1} 
\frac{ \left( \triangle_{m}^{2} h_i[k] \right)^2 }{2 (m \tau)^2},
\end{equation}
which is used to evaluate the resulting GTS in numerical examples below.

\subsection{Problem Statement}
The goal of this paper is to design the timing system to satisfy the two requirements in Section~\ref{sec:back}, which are interpreted mathematically as follows.

\textbf{Consistent Accuracy}: Synchronization of the MACs is said to be achieved if there exists a common state $x^*[k]\in \mathbb{R}^2$ such that
\begin{equation}\label{eq:obj_sync}
  \{x_i[k] - x^*[k]\}_{k\in\mathbb{Z}} \in \mathds{G}, \quad
\forall i \in \mathcal{N},
\end{equation}
which means that the mean of $x_i[k]$ is $x^*[k]$ and the covariance of the synchronization error is finite for all MACs at steady state.
In addition, to make the synchronized MACs have the least short-term AVAR, we address the optimization problem
\begin{equation}\label{eq:obj_reg_s}
\operatorname*{min}_{u_i[k]} \hat{\sigma}_{\sf A}^2 \bigl(\{h_i[k]\}, \tau \bigr).
\end{equation}
\textbf{Accuracy in Normal Operation}: 
Using GNSS signals, the MACs are anchored towards the standard time in the sense that
\begin{equation}\label{eq:obj_anchor}
\{x_i[k]\}_{k\in\mathbb{Z}} \in \mathds{G}, \quad \forall i\in \mathcal{N}.
\end{equation}
\textbf{Accuracy in Emergency Operation}: 
In the case of GNSS failures, the focus shifts to suppressing the drift rate of the MACs away from the standard time. 
We seek for optimal long-term floating control strategy by solving
\begin{equation}\label{eq:obj_reg_l}
\operatorname*{min}_{u_i[k]} \hat{\sigma}_{\sf A}^2 \bigl(\{h_i[k]\}, T \bigr),
\end{equation}
where $ T \gg \tau $ denotes the long-term control period.

\section{Lower-Level Distributed  Control in Hierarchical Atomic Timing Architecture}
\label{sec:Architecture1}
\subsection{State-Space Expansion}
\label{sec:State-expan}
In this section, we derive a decomposed system model based on state-space expansion to facilitate the subsequent control design.

The overall network topology is an undirected graph denoted by $\mathcal G=\{\mathcal{N},\mathcal E\}$,
where 
$\mathcal E \subseteq \mathcal{N} \times \mathcal{N}$ is the set of edges.
The adjacency matrix $\mathcal A=[a_{ij}] \in \mathbb R^{N \times N}$ describes the connections between MACs, where $a_{ij} = 1 $ if there is an edge connecting MACs $i$ and $j$, and $a_{ij} = 0 $ otherwise.
The neighborhood of MAC $i$ is denoted by $\mathcal{A}_i := \{j \in \mathcal{N}|(i,j) \in \mathcal E\}$.
Define $\mathcal{J}_i:=|\mathcal{A}_i|$, which is referred to as the degree of node $i$, and group them into $\mathcal{J} = \sfdiag(\mathcal{J}_1,\cdots,\mathcal{J}_N)$.
The Laplacian matrix of $\mathcal G$ is defined as $\mathcal L := \mathcal{J} - \mathcal{A}$.
Each undirected edge can be treated as two directed sub-edges with opposite directions, where the direction implies the source and sink of data flow. 
A sub-edge $g_{ij}$ is defined to describe the flow from MAC $j$ to MAC $i$, and the sub-edge set is denoted as $\mathcal E_i := \{g_{ij} | j \in \mathcal{A}_i\}$, which satisfies $|\mathcal E_i| = \mathcal{J}_i$.

For the stacked vectors $x[k]:= \sfcol (x_{i}[k])_{i \in \mathcal{N}}$, $y_i[k] := \sfcol (y_{ij}[k])_{j\in \mathcal{A}_i}$, and $w_i[k] := \sfcol (w_{ij}[k])_{j\in \mathcal{A}_i}$, we define the incidence matrix $V_i \in \mathbb R^{ \mathcal{J}_i \times N}$ as 
\blue{
\begin{equation}\label{eq:def_V}
V_i := \sfcol\bigl( \bm{\e}_{N|j}^{\sf T} - \bm{\e}_{N|i}^{\sf T} \bigr)_{j \in \mathcal{A}_i},
\end{equation}
where $\bm{\e}_{N|i}$ is the $i$th column of $I_N$.
}
Then, we have
\begin{equation}\label{eq:measure1}
y_i[k] = V_i (C \otimes I_N) x[k]+ w_i[k].
\end{equation}
Furthermore for $v[k]:= \sfcol (v_{i}[k])$, $u[k]:= \sfcol (u_{i}[k])$, $y[k]:= \sfcol (y_{i}[k])$, and $w[k]:= \sfcol (w_{i}[k])$, 
the overall model of the clock ensemble combining \eqref{eq:dmodel} and \eqref{eq:measure1} is summarized as
\begin{equation}\label{eq:ensmmld}
\simode{
x[k+1]&= \bm{A} x[k] + \bm{B} u[k] + v[k], \\
y[k] &= \bm{C} x[k] + w[k],
}
\end{equation}
where $\bm{A}:=  A \otimes I_N$, $\bm{B}:=  B \otimes I_N$, $\bm{C}:=  C \otimes V$, and $V := [V_1^{\sf T},\ldots,V_N^{\sf T}]^{\sf T}$.
Assuming that the noise of different MACs are independent, i.e., the covariance matrix of $v[k]$ is given by
\begin{equation}\label{eq:covQ}
\bm{Q}(\tau) = \mat{
\tau \Sigma_1 + \frac{\tau^3}{3} \Sigma_2   & \frac{\tau^2}{2}\Sigma_2  \\
\frac{\tau^2}{2} \Sigma_2 & \tau \Sigma_2 
}
\end{equation}
where $\Sigma_1:= \sfdiag(\sigma_{i1}^2)$ and $\Sigma_2:= \sfdiag(\sigma_{i2}^2)$. The covariance matrix of $w[k]$ is $\bm{R} := RI_{2 |\mathcal{E}|}$.

\blue{
As pointed out in \cite{ishizaki2026explicit}, clock ensemble models based solely on relative measurements inherently have an unobservable subspace, within which the ensemble GTS lies. 
In the centralized setting, this unobservable subspace can be explicitly isolated via observable canonical decomposition, and its properties can be manipulated using a deterministic Kalman filtering algorithm. 
We extend this idea to the distributed case as follows. 
}

Consider a normalized weighting vector $q\in\mathbb{R}^N$ as
\begin{equation}\label{eq:def_q}
q := \bigl(\mathds{1}_N^{\sf T}D^{-1}\mathds{1}_N \bigr)^{-1}D^{-1}\mathds{1}_N,
\end{equation}
where $D\in\mathbb{R}^{N\times N}$ is a positive definite diagonal matrix to be determined later. 
A skew projection matrix $\varPi \in\mathbb{R}^{N\times N}$ associated with $q$ 
 is defined as
\begin{equation}\label{eq:def_Pi}
\varPi:= I_N - \mathds{1}_N q^{\sf T},
\end{equation}
which satisfies $\sfim \varPi = \sfker q^{\sf T}$ and $\sfker \varPi = \sfim \mathds{1}_N$.
Using these matrices, we define the augmented state as
\begin{equation}\label{eq:def_sysdecom}
\mat{z[k] \\ \bar{z}[k]} := \mat{  I_2 \otimes \varPi \\ I_2 \otimes q^{\sf T} } x[k].
\end{equation}
Consequently, an equivalent state-space expansion associated with $q$ is given as
\begin{equation}\label{eq:state_expan}
x[k] = z[k] + (I_2 \otimes \mathds{1}_N) \bar{z}[k] .
\end{equation}
The control input can be expanded in the same way as
\begin{equation}\label{eq:control_expan}
u[k] = \eta[k] + (I_2 \otimes \mathds{1}_N) \bar{\eta}[k] .
\end{equation}
Then, an expanded model of \eqref{eq:ensmmld} is obtained as
\begin{align}
\simode{ \label{eq:dyn_z}
z[k+1] &= \bm{A} z[k] + \bm{B} \eta[k] + (I_2 \otimes \varPi) v[k], \\
y[k] & = \bm{C} z[k] + w[k],
}\\
\bar{z}[k+1] = A \bar{z}[k] + B \bar{\eta}[k] + (I_2 \otimes q^{\sf T}) v[k]. \label{eq:dyn_zbar}
\end{align}
In this augmented system, $\bar{z}[k]\in \mathbb{R}^{2}$ equals to the projection of $x[k]$ onto $\sfim (I_2 \otimes q^{\sf T})$, which represents a weighted average of all states. 
On the other hand, $z[k] \in \mathbb{R}^{2N}$ represents the deviations of $x[k]$ from the common center $\bar{z}[k]$.
This state-space expansion is consistent with the goals of atomic timing. 
In particular, stabilizing $z[k]$ leads to synchronization for the clock ensemble, while stabilizing $\bar{z}[k]$ improves the accuracy. 

Note that the subsystem \eqref{eq:dyn_zbar} is not observable from the relative measurement $y[k]$. 
To estimate $\bar{z}[k]$, we must use external GNSS anchors as a standard time reference and measure the time difference from them. 
However, if the anchors are not connected, the goal must be to reduce the drift of the system from the ideal time coordinate. 
The following sections provide technical details for each component of the timing system.
\subsection{GNSS-Free Distributed Synchronization}
\label{sec:Sync}
In this section, we design a distributed synchronization control strategy that uses only local information exchange between adjacent MACs. 
This strategy is independent of GNSS and is commonly used in both normal and emergency operating modes.

We introduce the notion of an ``edge state", defined as
\begin{equation}\label{eq:def_edge_state}
\zeta_{ij}[k] := x_j[k] - x_i[k]
\end{equation}
to quantify the state difference between two adjacent MACs.
Note that $\zeta_{ij}[k]$ and the sub-edge $g_{ij}$ from the MAC $j$ to the MAC $i$ have a one-to-one correspondence. 
Therefore, for an undirected graph, a pair of edge states, i.e., $\zeta_{ij}[k]$ and $\zeta_{ji}[k]$, are defined for a single edge. 
Denote the collection of the edge states of MAC $i$ by
\[
\zeta_{i}[k] := (I_2 \otimes V_i) x[k].
\]
Using \eqref{eq:state_expan} with $V_i \mathds{1}_N = 0$, we see that it is equivalent to
\begin{equation}\label{eq:defzetak}
\zeta_{i}[k] = (I_2 \otimes V_i) z[k].
\end{equation}
Thus, \eqref{eq:dyn_z} leads to the edge state dynamics 
\begin{equation}\label{eq:den_edge}
\simode{
\zeta_{i}[k+1] &= \bm{A}_i \zeta_i[k] + \bm{B}_i \breve{\eta}_i[k] + (I_2 \otimes V_{i}) v[k] \\
y_i[k] & = \bm{C}_i \zeta_{i}[k] + w_{i}[k],
}
\end{equation}
where the system matrices are denoted as $\bm{A}_i:= A \otimes I_{\mathcal{J}_i}$, $\bm{B}_i := B \otimes I_{\mathcal{J}_i}$, $\bm{C}_i := C \otimes I_{\mathcal{J}_i}$, and $\breve{\eta}_i[k] := (I_2 \otimes V_i) \eta[k]$.
Note that this system is observable because the pair $(A,C)$ is observable.
Using the measurements $y_i[k]$, we can develop a state estimator for each MAC by applying the Kalman filtering algorithm \cite{anderson2005optimal} in such a way that 
\begin{equation}\label{eq:den_edge_est}
\hat{\zeta}_i[k+1] = \bm{A}_i \hat{\zeta}_i[k]
+ \bm{B}_i \breve{\eta}_i[k]
+ \!\bm{H}_i \bigl( y_i[k] - \bm{C}_i \hat{\zeta}_i[k] \bigr).
\end{equation}
Since the timing lasts for an extended period of time, we can assume that the Kalman filter is in the steady state with the constant gain 
\[
\bm{H}_i = \mathcal{F}(\bm{A}_i,\bm{C}_i,(I_2 \otimes V_i)\bm{Q}(\tau)(I_2 \otimes V_i^{\sf T}), R \otimes I_{\mathcal{J}_i}),
\]
where $\mathcal{F}(A,C,Q,R)$ denotes the steady-state equations
\begin{equation}\label{eq:kalfil}
\!\!\!\simode{
&P = A P A\!^{\sf T}\! - \!A P C^{\sf T}(C P C^{\sf T} + R)^{-1} C^{\sf T} P A + Q,\\
&H = A P C^{\sf T} \bigl( C P C^{\sf T} + R \bigr)^{-1}.
}
\end{equation}
By such a decentralized estimator, it is guaranteed that the estimation error $e_i[k]:=\zeta_i[k] - \hat{\zeta}_i[k]$ is convergent in the sense that 
\begin{equation}\label{eq:sync_est_error}
\{e_i[k]\}_{k\in\mathbb{Z}} \in \mathds{G}, \quad \forall i\in \mathcal{N}.
\end{equation}
Based on the edge state estimation, we apply the distributed control strategy
\begin{equation}\label{eq:sync_con}
u_i [k] = D_i F_{s} \sum_{\blue{ j \in \mathcal{A}_i} } \hat{\zeta}_{ij}^+[k]
,\quad
i\in\mathcal{N},
\end{equation}
where $F_{s} \in \mathbb{R}^{1\times 2} $ is the control gain to be designed, $D_i$ is the $i$th diagonal element of $D$ in \eqref{eq:def_q}, and $\hat{\zeta}_{ij}^+[k]$ is an unbiased estimate for MACs $i$ and $j$, computed as
\begin{equation}\label{eq:est_fusion}
\hat{\zeta}^+_{ij}[k] = \hat{\zeta}_{ij}[k] - \frac{\hat{\zeta}_{ij}[k]+ \hat{\zeta}_{ji}[k]}{2}.
\end{equation}
Although, computing \eqref{eq:est_fusion} requires additional information exchange between neighboring MACs, this does not violate the distributed design principle. 
It should be emphasized that due to this unbiased estimation, it will be shown in Theorem~\ref{theo:sync} that the unobservable synchronization destination $\bar{z}[k]$ is independent of the distributed synchronization control.

For the following discussion, we define a particular free-running dynamics as
\begin{equation}\label{eq:syncdes}
\Phi(q): 
\simode{
r[k+1] &= A r[k] + (I_2 \otimes q^{\sf T}) v[k], \\
h_r[k] &= C r[k],
}
\end{equation}
which is the weighted average of the uncontrolled MACs associated with $q$.
Then, \textbf{Requirement \uppercase\expandafter{\romannumeral1}} is satisfied as follows. 

\begin{theorem}\label{theo:sync}
For the system \eqref{eq:ensmmld}, consider the decentralized estimator \eqref{eq:den_edge_est} and distributed control strategy \eqref{eq:sync_con}. 
Suppose that the communication network is connected, and the feedback gain $F_{s}$ is structured as
\begin{equation}\label{eq:F_s}
    F_{s} = \gamma_{s} \mat{ \frac{\alpha_{s}}{\tau} ~&~ 1 },
\end{equation}
where $\gamma_{s}$ and $\alpha_{s}$ are scalar parameters.
Then 
\begin{equation}\label{eq:delta_s}
    \{x_i[k] -  r[k]\}_{k\in\mathbb{Z}} \in \mathds{G}, \quad \forall i \in \mathcal{N}
\end{equation}
for the free-running dynamics $\Phi(q)$ in \eqref{eq:syncdes} if and only if
\begin{equation}\label{eq:con_Fz}
0 < \gamma_{s} < \tfrac{4}{2 + \alpha_{s}} \lambda_{\rm max}^{-1}(D^{\frac{1}{2}} \mathcal L D^{\frac{1}{2}}), \quad \alpha_{s} >0,
\end{equation}
where $\lambda_{\rm max}(\cdot)$ denotes the largest eigenvalue.
\end{theorem}

\begin{proof}
The distributed control strategy in \eqref{eq:sync_con} can be expressed as
\[
u[k] = (F_{s} \otimes DS) (\zeta[k] + \Delta e[k]),
\]
where $
S := \sfdiag
\bigl( \mathds{1}_{\mathcal{J}_i}^{\sf T} \bigr)_{i\in\mathcal{N}} \in \mathbb{R}^{N \times 2 |\mathcal{E}|} 
$ and $\Delta e[k]$ is the stacked vector of $\Delta e_{ij}[k] := (e_{ji}[k] - e_{ij}[k])/2$. 
Note that $S V = - \mathcal L$.
Therefore, it gives that
\[
u[k] = -(F_{s} \otimes D \mathcal L) z[k] + (F_{s} \otimes DS) \Delta e[k].
\]
Based on the expansion \eqref{eq:control_expan} and the definition \eqref{eq:def_q}, the multiplication of $(I_2 \otimes q^{\sf T})$ leads to
\begin{equation}\label{eq:eta_bar_0}
\bar{\eta}[k] = (F_{s} \otimes q^{\sf T} D S) \Delta e[k] = 0, \quad \forall k \in \mathbb{Z},
\end{equation}
because of $ \mathds{1}_N^{\sf T} \mathcal L = 0$ and $\sum_{i,j \in \mathcal{N}} \Delta e_{ij}[k] = 0$.
By comparing \eqref{eq:dyn_zbar} and \eqref{eq:syncdes}, we have $x_i[k] -r[k] = z_i[k]$. Thus, we only need to prove that $z_i[k]$ satisfies \eqref{eq:delta_s}.
The controlled dynamics of $z[k]$ in \eqref{eq:dyn_z} equals to
\[
z[k+1] = \bm{A}_s z[k] + (BF_{s} \otimes DS) e[k]  + (I_2 \otimes \varPi) v[k],
\]
where $\bm{A}_s = A \otimes I_N - BF_{s} \otimes D  \mathcal L$.
The eigenvalues of $\bm{A}_s$ are found as
\[
\bm{\Lambda}(\bm{A}_s)  = 
\bigcup_{\lambda \in \bm{\Lambda}(D^{\frac{1}{2}} \mathcal L D^{\frac{1}{2}})} \bm{\Lambda}(A-\lambda BF_{s}).
\]
Note that the characteristic polynomial of $A-\lambda BF_{s}$ is
\[
\sfdet \bigl\{zI_2- (A-\lambda BF_{s}) \bigr\} = z^2 -2 \left(1-\lambda \gamma_{s} \tfrac{1 + \alpha_{s}}{2} \right) z + 1-\lambda \gamma_{s}.
\]
\blue{
Based on the Jury stability test, its roots exist inside the open unit circle if and only if
\[
|2-\lambda\gamma_{s}(1+\alpha_{s})|-1<1-\lambda\gamma_{s} < 1.
\]
}
Since the topology is connected, the eigenvalues of $D\mathcal L$ are all positive expect one zero eigenvalue.
Therefore, the condition \eqref{eq:con_Fz} ensures that $\bm{A}_s$ is marginally Schur-stable.
According to \cite{olfati2004consensus} and the results in \eqref{eq:sync_est_error}, $z[k]$ is convergent in the sense that
\[
\lim_{k \to \infty} \mathbb E[z[k]] \in \sfim(I_2 \otimes q^{\sf T}).
\]
On the other hand, notice that $z[k] \in \sfim(I_2 \otimes \varPi)$, which is equal to $\sfker(I_2 \otimes q^{\sf T})$, leads to $\lim_{ k \to \infty} \mathbb{E}[z[k]] = 0$. 
Furthermore, since both $e[k]$ and $v[k]$ have finite covariances, it is easy to conclude that the covariance of $z[k]$ is also bounded. 
This completes the proof.
\hfill $\square$
\end{proof}

In addition to synchronism, the next goal is to maximize the frequency stability of the synchronized time, or GTS, as measured by AVAR.
Theorem~\ref{theo:sync} shows that under the distributed control strategy \eqref{eq:sync_con}, all MACs converge to the specific free-running dynamics $\Phi(q)$, which can be controlled by choosing the free parameter $q$.
Its AVAR with sampling period $ m \tau$ can be analytically derived as
\begin{align}\label{eq:allan_ANA}
\sigma_{\sf A}^2 \bigl(m \tau \bigr)  &=
\frac{G\!
\left\{
I_2 \otimes\! \left[ (I_2 \otimes q^{\sf T})\bm{Q}(m \tau) (I_2 \otimes q) \right]
\right\}\!
G^{\sf T}}{2(m \tau)^2} \nonumber \\
&=
\frac{q^{\sf T}  \varGamma(m \tau) q}{(m \tau)^2},
\end{align}
where $G := [ C(A-2 I_2)~C ]$ and 
\begin{equation}
    \varGamma(m \tau) := (m \tau) \Sigma_1 + \frac{(m \tau)^3}{3} \Sigma_2.
\end{equation}
Based on this relationship, the optimal $q$ that leads to the least AVAR is found as follows.
\begin{corollary}
\label{theo:reg_short}
Consider the system \eqref{eq:ensmmld} under the synchronization strategy in Theorem \ref{theo:sync}. 
Then the optimal weighting vector 
\begin{equation}\label{eq:lstAV_arg}
q_{\sf A}(\tau) :=  \arg \min_{q} \sigma_{\sf A}^2 \bigl(\tau \bigr)
\quad {\rm s.t.} \quad
q^{\sf T} \mathds{1}_N =1,
\end{equation}
which minimizes the AVAR over $\tau$, is given as
\begin{equation}\label{eq:lstAV}
q_{\sf A}(\tau) = 
\frac{\varGamma^{-1}(\tau) \mathds{1}_N}{\mathds{1}_N^{\sf T} \varGamma^{-1}(\tau) \mathds{1}_N }.
\end{equation}
\end{corollary}
\begin{proof}
Since it has been proved that $x[k]$ converges to $(I_2 \otimes \mathds{1}_N^{\sf T}) r[k]$ in the sense of expectation, solving \eqref{eq:obj_reg_s} is equivalent to solving \eqref{eq:lstAV_arg}.
\blue{
Define the Lagrangian 
\[
L(q,\lambda) = \frac{1}{\tau^2} q^{\sf T} \Gamma(\tau) q - \lambda \bigl(q^{\sf T} \mathds{1}_N - 1\bigr).
\]
Taking its gradient with respect to $q$ and letting the gradient equal to zero yields that the optimum is found as the solution to
\[
2\tau^{-2}\varGamma(\tau) q -\lambda \mathds{1}_N =0
,\quad
q^{\sf T} \mathds{1}_N =1,
\]
}
which is given as $\eqref{eq:lstAV}$. This proves the claim.
\hfill $\square$
\end{proof}
It is shown in \eqref{eq:lstAV} that the optimal weighting vector depends on a combination of the noise, which is an implicit function of the sampling interval $m\tau$. 
For standard MACs, $\Sigma_{1}$ is several orders of magnitude larger than $\Sigma_{2}$. 
Thus, two representative results can be obtained as
\begin{align}
&q_0:= \lim_{m \to 0} q_{\sf A}(m\tau) = 
\frac{\Sigma_1^{-1} \mathds{1}_N}{\mathds{1}_N^{\sf T} \Sigma_1^{-1} \mathds{1}_N }, \label{eq:q_zero}\\
&q_\infty:= \lim_{m \to \infty} q_{\sf A}(m\tau) = 
\frac{\Sigma_2^{-1} \mathds{1}_N}{\mathds{1}_N^{\sf T} \Sigma_2^{-1} \mathds{1}_N }. \label{eq:q_inf}
\end{align}
In fact, $q_0$ is the weighting vector for the best short-term frequency stability, while $q_\infty$ is for the best long-term frequency stability.
These two weighting vectors can be used to evaluate the limits of system performance in practice.

\section{Upper-Level Broadcast Control in Hierarchical Atomic Timing Architecture}
\label{sec:Architecture2}

\subsection{GNSS-Based Anchoring in Normal Operation} 
\label{sec:Anchoring}
In the normal operating mode, the MACs adjacent to the GNSS anchors periodically receive the standard time with the period $T$. 
Assume that $T$ is an integer multiple of $\tau$, i.e., $\ell := T / \tau$ is an integer.
In the following analysis, the new time index $K$ is used to represent the time instant $K T$.

The edge states between the anchors and their adjacent MACs are defined as
\[
Z_{li}[K] := X_l[K] - x_i[K],\quad l \in \mathcal{M}, \quad i \in \mathcal{W}_l.
\]
Using the measurement $Y_{li}[K]$, a state estimator can be constructed as
\[
\hat{Z}_{li}[K\!+\!1] = \bm{A}_a \hat{Z}_{li}[K]
+ \bm{U}^a_{li}[K]
+ \!H^a_{li} \bigl( Y_{li}[K] \!-\! C \hat{Z}_{li}[K] \bigr),
\]
where $\bm{A}_a := A^{\ell}$, and
\begin{align*}
&\bm{U}^a_{li}[K] := \sum_{k=0}^{\ell - 1}\! A^{ \ell - 1 - k} B u_i[K \ell + k],\\
& H^a_{li} = \mathcal{F}\left(\bm{A}_a,C,Q^a_l(T)+Q_i(T),R \right),
\end{align*}
which is updated intermittently. 
Since the ensemble has been guaranteed to keep synchronized, each $\hat{Z}_{li}[K]$ can also be regarded as a separate estimate of the same $\bar{z}[K]$.
Therefore, we apply the anchoring strategy
\begin{equation}\label{eq:control_anchor}
\bar{\eta}[k]:= \simode{
- F_{a} & W \hat{Z}[K], && k=K \ell, ~K \in \mathbb Z, \\
&0, &&\text{otherwise},
}
\end{equation} 
where $F_a \in \mathbb{R}^{1 \times 2}$ is the control gain to be designed, $W := I_2 \otimes \mathds{1}_{|\mathcal{W}|}^{\sf T}/|\mathcal{W}|$ is an averaging vector, and \[
\hat{Z}[K]:= \sfcol\{\sfcol\{\hat{Z}_{li}[K]\}_{i\in\mathcal{W}_l}\}_{l\in\mathcal{M}}.
\]
This anchoring strategy is common to all MACs.  
Here we directly use the notation $\bar{\eta}[k]$ to indicate that such common control has no effect on the synchronization, due to the fact that $\bar{\eta}[k] \in \sfker(I_2 \otimes \varPi)$.
Then, \textbf{Requirement \uppercase\expandafter{\romannumeral2}} in the normal operating mode is satisfied as follows. 

\begin{theorem} \label{theo:anchor}
Consider the system \eqref{eq:ensmmld} under synchronization algorithm in Theorem \ref{theo:sync} and anchoring strategy in \eqref{eq:control_anchor}. 
For a given $T >0$, suppose that the feedback gain $F_{a}$ is structured as 
\begin{equation}\label{eq:F_a}
    F_{a} = \gamma_{a} \mat{ \frac{\alpha_{a}}{T} ~&~ 1 },
\end{equation}
where $\gamma_{a}$ and $\alpha_{a}$ are scalar parameters.
Then 
\[
\{ x[k] - (\theta^* e_{2|1} \otimes \mathds{1}_N)\}_{k \in \mathbb{Z}} \in \mathds{G}
\]
for any initial condition, where $\theta^*:= \sum_{l\in\mathcal{M}} |\mathcal{W}_l|\Theta_l^* / |\mathcal{W}|$, if and only if 
\begin{equation}\label{eq:con_Fa}
0< \gamma_{a} <\frac{4}{2+\alpha_{a}}, \quad
\alpha_{a} >0.
\end{equation}
\end{theorem}

\begin{figure}[t]
\centering
\includegraphics[scale=0.4]{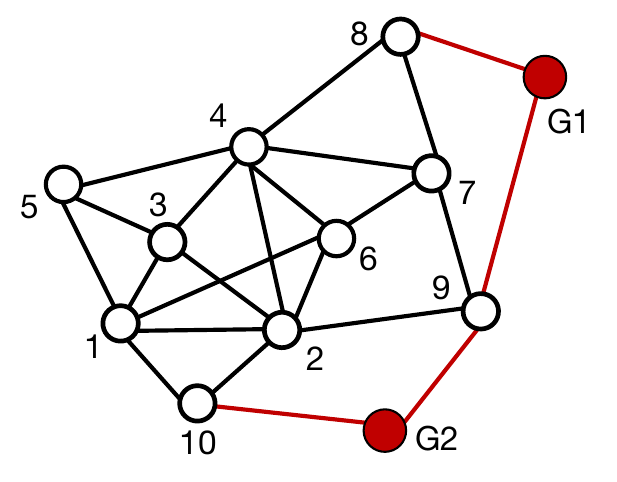}
\caption{The network of 10 MACs and 2 anchors.}
\label{fig:topology}
\end{figure}

\begin{proof}
The edge state can be rewritten as
\[
Z[K] = z[K] + (I_2 \otimes \mathds{1}_N) \bar{z}[K] - X[K],
\]
where $Z[K]$ is defined similarly to $\hat{Z}[K]$. 
Denote the estimation error as $E[K]:=Z[K] - \hat{Z}[K]$. 
Then, because of \eqref{eq:Exanc}, we have
\[
W \hat{Z}[K] = (\bar{z}[K] - \theta^* e_{2|1}) + W(z[K] + E[K]).
\]
For the deviation $\Delta \bar{z}[K] := \bar{z}[K] - \theta^* e_{2|1}$, we know
\begin{align*}
\Delta \bar{z}[K+1] 
=& A^{\ell} \Delta \bar{z}[K] + A^{\ell-1} B \bar{\eta}[K] + (I_2 \otimes q^{\sf T}) V^a[K] \\
=& (A^{\ell} - A^{\ell-1} B F_{a}) \Delta \bar{z}[K] + (I_2 \otimes q^{\sf T}) V^a[K] \\
& ~~~~ - B F_{a} W(z[K] + E[K]),
\end{align*}
where $V^a[K] = \sum_{k=0}^{\ell-1} A^{\ell-1 - k} v[K \ell + k]$ and $A^{\ell} e_{2|1} = e_{2|1}$ for any $\ell$. 
The roots of the characteristic polynomial of $(A^{\ell} - A^{\ell-1} B F_{a})$ exist inside the open unit circle if and only if
$|2-\gamma_{a}(1+\alpha_{a})|-1<1-\gamma_{a} < 1$.
Therefore, the condition \eqref{eq:con_Fa} ensures that this system is internally stable.
Moreover, since it has been guaranteed that $z[k]$ and $E[k]$ are convergent, it follows that $\Delta \bar{z}[k]$ is stabilized. This completes the proof.
\hfill $\square$
\end{proof}
\textbf{Illustrative Example}:
A numerical example is given here to demonstrate the proposed strategies in the normal operating mode.  
Consider a network consisting of 10 MACs and 2 anchors. 
The network structure is shown in Fig.~\ref{fig:topology}, where the MACs are labeled from 1 to 10 and the anchors are labeled by G1 and G2.
Each MAC is written by \eqref{eq:dmodel} with the sampling period $\tau = 1~\text{[sec]}$. 
The variances of white frequency noise $\sigma_1^2$ and random walk frequency noise $\sigma_2^2$ are listed in Table \ref{table:noisepara}, whose values are determined based on our experiments. The variance of the measurement noise is set to $10^{-24}$. The long-term control period is set to $T = 2000~\text{[sec]}$.

\begin{table*}[t]
\caption{Variances of white frequency noise and random walk frequency noise.}
\label{table:noisepara}
\centering
\begin{tabular}{lccccccccccccccc}
& scale & 1  &  2 &   3  &  4  &  5  &  6  &  7  &  8  &  9  &  10  \\
\hline
$\sigma_1^2$ & $ 10^{-20}$ & 3.31  &  0.887 &   1.51  &  1.93  &  9.33  &  1.31  &  3.87  &  5.26  &  0.981  &  3.39  \\
$\sigma_2^2$ &  $ 10^{-26}$ &  3.12  &  0.295  &  1.52  &  6.97  &  7.74  &  0.251  &  0.106  &  0.765  &  0.207  &  0.38  \\  \hline
\end{tabular}
\end{table*}

\begin{figure}[t]
\centering
\includegraphics[scale=0.38]{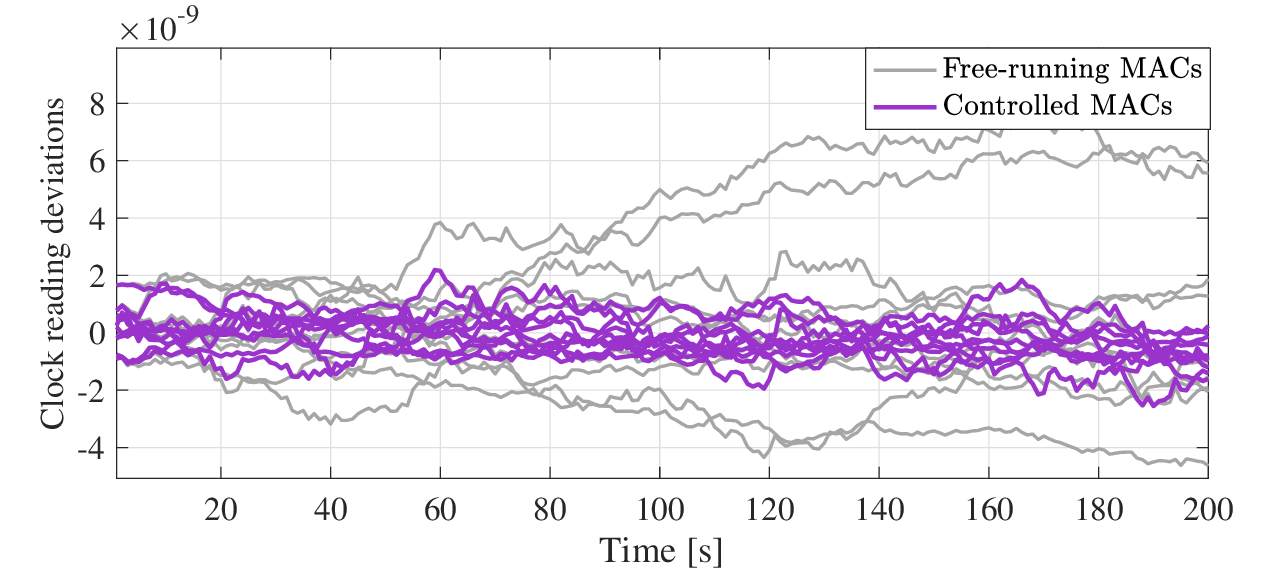}
\caption{Clock reading deviations under the distributed control strategy.}
\label{fig:State_1e2}
\end{figure}

The behavior of the MACs under the proposed distributed synchronization strategy is shown in Fig.~\ref{fig:State_1e2}. 
The clock reading deviations of the free-running MACs, i.e., $ h_{i}[k]$ for $i \in \mathcal{N}$, are plotted as the gray lines. 
They start with different initial values, and continue to diverge due to the non-zero frequency differences.
On the other hand, as shown in Theorem~\ref{theo:sync}, the controlled clocks achieve synchronization, plotted as the purple lines in Fig.~\ref{fig:State_1e2}. 

\begin{figure}[t]
\centering
\includegraphics[scale=0.38]{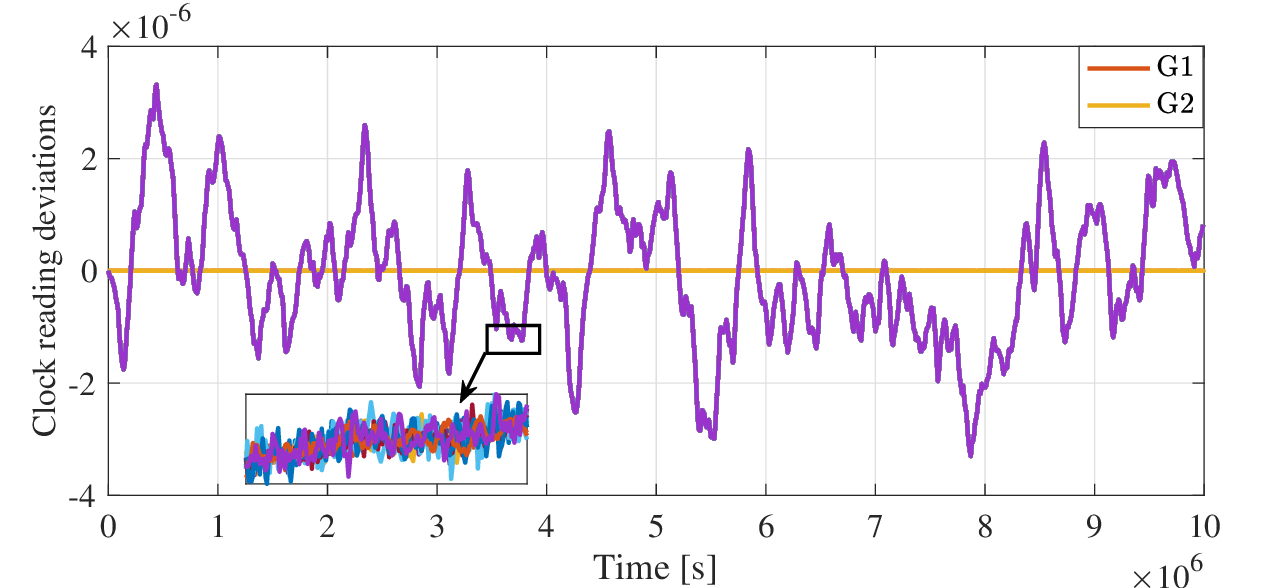}
\caption{Clock reading deviations anchored by GNSS signals in the normal operating mode.}
\label{fig:State_1e7}
\end{figure}

Next, we examine the long-term frequency stability of the GTS anchored by GNSS in the normal operating mode. 
The clock reading deviations for the time period $k \in [1,10^7]$ are shown in Fig.~\ref{fig:State_1e7}, where those of the anchors are also plotted. Even with a long control period, GTS stays close to the standard time by anchoring.
Note also that the MACs are always synchronized. 
For reference, the MACs without anchoring will gradually diverge up to the level of $10^{-4}$ at $k=10^7$. 
This clearly illustrates the validity of our anchoring strategy.

\begin{figure}[t]
\centering
\includegraphics[scale=0.4]{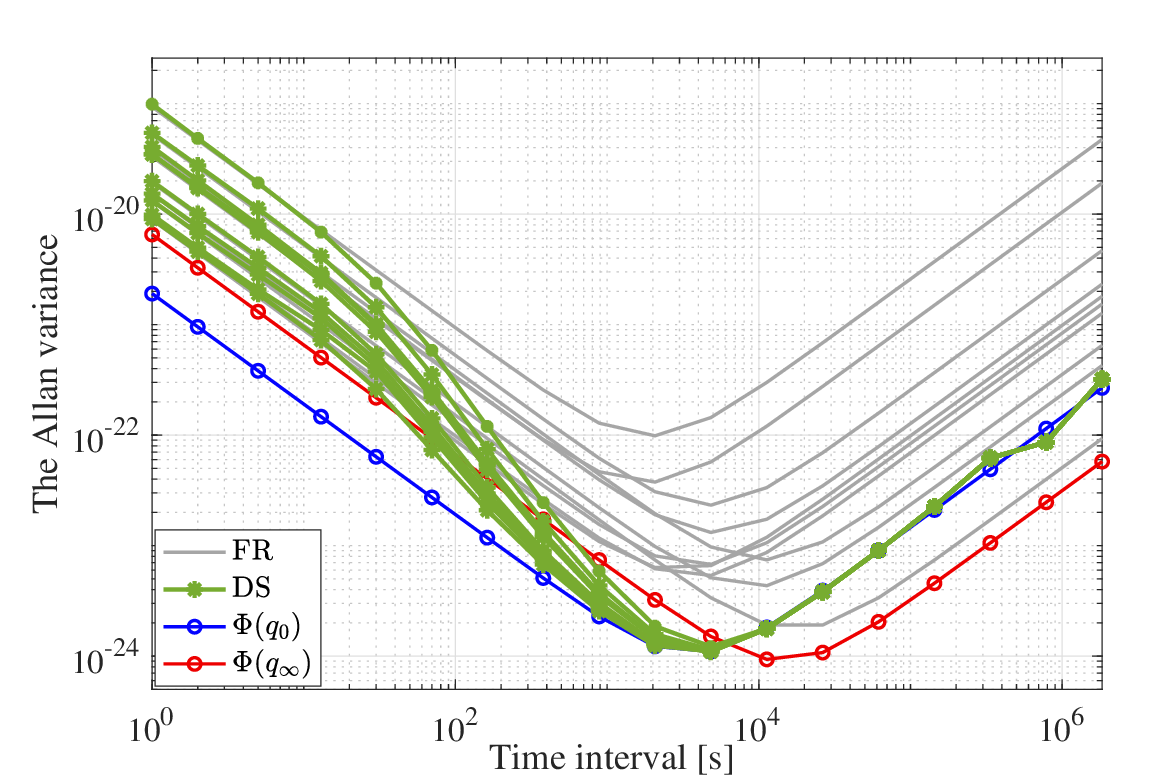}
\caption{The AVAR of the MACs that are free-running (FR), controlled by the distributed synchronization strategy alone (DS), and the free-running ensemble mean $\Phi(q_{0})$ and $\Phi(q_{\infty})$.}
\label{fig:sync_avar}
\end{figure}

To illustrate the need for control mode switching during GNSS failures in Section~\ref{sec:Floating} below, we show the AVAR curves of the MACs without GNSS anchoring in Fig.~\ref{fig:sync_avar}. 
\blue{
The AVAR of the synchronized MACs are evaluated based on the simulated time series using the statistical estimator \eqref{eq:allan_est}, whereas those of the free-running MACs are calculated a priori using the analytical
formula \eqref{eq:allan_ana}. 
}
This figure also shows the short- and long-term optimal AVAR curves for the free-running MACs, i.e., $\Phi(q_{0})$, and $\Phi(q_{\infty})$, computed according to \eqref{eq:allan_ANA}, as the blue and red lines.
We can see that the synchronization control can maximize the short-term stability, but fail for longer periods when GNSS is unavailable.
This motivates us to develop a different upper-level control strategy to maximize long-term frequency stability in emergency operation without GNSS anchors.

\subsection{Optimal Floating in Emergency Operation}
\label{sec:Floating}

In this section, we propose an optimal control strategy for the GTS in case of GNSS failures. 
In particular, in addition to \eqref{eq:obj_reg_s}, we also aim to maximize the long-term frequency stability as in \eqref{eq:obj_reg_l}.

Note that $\bar{z}[k]$ is not observable due to the lack of an external standard time reference.
To overcome this difficulty, we next construct an estimator capable of inferring the deviation of $\bar{z}[k]$ from the desired optimum. 
In particular, the following analysis is built on the basis of $q_{\tau}$ and $q_{T}$, which are shorthand for $q_{\sf A}(\tau)$ and $q_{\sf A}(T)$, respectively. 
Unless otherwise specified, variables associated with $q_{\tau}$ or $q_{T}$ are denoted by their subscripts. 

The system states associated with $q_{\tau}$ and $q_{T}$ can be expanded as 
\[
\!\mat{z_{\tau}[k]\! \\ \!\bar{z}_{\tau}[k]\!} = \mat{  I_2 \otimes \varPi_{\tau}\! \\ I_2 \otimes q_{\tau}^{\sf T}\! } x[k], ~ \mat{  \!z_{T}[k]\! \\ \!\bar{z}_{T}[k]\!} = \mat{  I_2 \otimes \varPi_{T}\! \\ I_2 \otimes q_{T}^{\sf T}\! } x[k].
\]
Combing with \eqref{eq:state_expan} and $\varPi_{\tau} \mathds{1}_N = 0$, we have
\[
\mat{
z_{\tau}[k] \\
\bar z_{\tau}[k]
}
=
\mat{
I_2 \otimes \varPi_{\tau} & 0 \\
I_2 \otimes q_{\tau}^{\sf T} & I_2
}
\mat{
z_{T}[k] \\
\bar z_{T}[k]
}.
\]
This shows that the difference between $\bar z_{\tau}[k]$ and $\bar z_{T}[k]$ is related to the synchronization error. 
\blue{
In \cite{ishizaki2026explicit}, a determinate Kalman filter is developed to estimate this difference for regulating the synchronization destination. 
Unlike in the centralized case, the error itself remains difficult to evaluate in a distributed manner. 
However, since $\zeta[k]$ contains the information of $z[k]$, it can serve as an intermediary. 
Next we seek a minimal necessary subset that transforms $\zeta[k]$ back into $z[k]$. 
}

The undirected network topology among MACs must contain a spanning tree. 
By assigning an arbitrary direction to each edge of the spanning tree, we can find a directed tree, denoted by $\mathcal G_{\beta}$, and the rest is denoted by $\mathcal G_{\bar \beta}$.
Let $\zeta_{\beta}[k]$ denotes the vector composed of all edge states $\zeta_{ij}[k]$ corresponding to each $g_{ij} \in \mathcal E(\mathcal G_{\beta})$.
Similarly, denote the collection of all $\zeta_{ij}[k]$ associated with $g_{ij} \in \mathcal E(\mathcal G_{\bar \beta})$ by $\zeta_{\bar \beta}[k]$.
For each directed edge $g_{ij} \in \mathcal E (\mathcal G_{\bar \beta})$ from MAC $i$ to MAC $j$, there always exists an equivalent directed path within $\mathcal G_{\beta}$ that starts from $i$ up to the root node and then descends to $j$. 
Therefore, there exists a matrix $T_{\bar \beta} \in \mathbb R^{(|\mathcal{E}|-N+1) \times (N-1)}$ such that
\begin{equation}\label{eq:zeta_bar_beta}
\zeta_{\bar \beta}[k] = (I_2 \otimes T_{\bar \beta}) \zeta_{\beta}[k].
\end{equation}
Denote the incidence matrix of $\mathcal G_{\beta}$ by $V_{\beta} \in \mathbb R^{(N-1) \times N}$, and that of $\mathcal G_{\bar \beta}$ by $V_{\bar \beta}$. 
Then, we have the following lemma.
\blue{
\begin{lemma}
\label{lmm:trans_z}
Let $q$ be a given weighting vector, and let $W \in \mathbb{R}^{N\times (N-1)}$ be any full-column-rank matrix satisfying $q^{\sf T} W = 0$. 
For any directed spanning tree with associated $V_{\beta}$, the transformation
\begin{equation}\label{eq:def_zetam}
z[k] = (I_2 \otimes V_{\beta}^+) \zeta_{\beta}[k] 
\end{equation}
holds, where $\zeta_{\beta}[k] \in \mathbb R^{2(N-1)}$ is the corresponding edge state vector, and $V_{\beta}^+ \in \mathbb R^{N \times (N-1)}$ is the generalized inverse of $V_{\beta}$ induced by $q$, defined as
\begin{equation}\label{eq:def_geninv}
V_{\beta}^+ := W (V_{\beta} W)^{-1}.
\end{equation}
\end{lemma}
\begin{proof}
From \eqref{eq:defzetak} and \eqref{eq:zeta_bar_beta}, we know
\[
\left( I_2 \otimes \mat{V_{\beta} \\ V_{\bar \beta}} \right) z[k] = \left( I_2 \otimes \mat{I_{N-1} \\ T_{\bar \beta} } \right) \zeta_{\beta}[k].
\]
Because $V_{\beta}$ has rank $N-1$ and satisfies $V_{\beta}\mathds{1}_N=0$, the inverse mapping from $\zeta_{\beta}[k]$ to $z[k]$ is not unique. 
By \eqref{eq:def_sysdecom}, $z[k]$ must satisfy $z[k] \in \ker(I_2\otimes q^{\sf T})$. 
Therefore, to enforce this, $W$ is chosen so that $\sfim(W)=\sfker(q^{\sf T})$, and the generalized inverse in \eqref{eq:def_geninv} yields $\sfim(V_{\beta}^+) = \sfker(q^{\sf T})$. 
This claims the result.
\hfill $\square$
\end{proof}
}
\blue{
\textbf{Remark }
The network topology may contain multiple distinct directed spanning trees, each consisting of $N-1$ edges that connect all $N$ MACs. 
Theoretically, any such tree suffices to support the functioning of the optimal floating control. 
In practice, however, it is preferable to select a tree whose edges correspond to communication links with higher reliability, lower latency, and a lower average measurement noise level.
}

Using Lemma \ref{lmm:trans_z} for $z_{\tau}[k]$ and $z_{T}[k]$ with $\zeta_{\beta}[k]$, we have
\begin{equation}\label{eq:basisqinfq}
\mat{\zeta_{\beta}[k] \\ \bar z_{\tau}[k]} 
=\mat{
I_2 \otimes I_{N-1}  & ~0 \\
I_2 \otimes q_{\tau}^{\sf T} V_{\beta T}^+ & ~I_2
} \mat{\zeta_{\beta}[k] \\ \bar z_{T}[k]},
\end{equation}
\blue{
where $V_{\beta T}^+$ denotes the generalized inverse of $V_{\beta}$ associated with $q_{T}$.
}
We use the same symbol $\zeta_{\beta}[k]$ on both sides, as it denotes the same quantity.
The relation in \eqref{eq:basisqinfq} provides a key insight that the gap between $\bar z_{\tau}[k]$ and $\bar z_{T}[k]$ can be filled by $\zeta_{\beta}[k]$, denoted as $\bar{z}^-_{T}[k] :=(I_2 \otimes q_{\tau}^{\sf T} V_{\beta T}^+) \zeta_{\beta}[k]$. 
Thus, if the current GTS is $\bar z_{\tau}[k]$, the GTS will shift from $\bar z_{\tau}[k]$ to $\bar z_{T}[k]$ by drowning out $\bar{z}^-_{T}[k]$.

Based on this observation, an state estimator is given as
\begin{align}
&\simode{\label{eq:est_zeta_beta}
\hat{\zeta}_{\beta}[k+1] & = \bm{A}_{f} \hat{\zeta}_{\beta}[k] + \bm{B}_{f} \breve{\eta}_{\beta}[k]
+ \bm{H}_{f} \Delta y_{\beta}[k],\\
\Delta y_{\beta}[k] &= y_{\beta}[k] - \bm{C}_{f} \hat{\zeta}_{\beta}[k],
}\\
&~~~~\hat{\bar{z}}^-_{T}[k+1] = A \hat{\bar{z}}^-_{T}[k] + B \bar{\eta}[k] 
+ \bm{H}_{\bar{f}} \Delta y_{\beta}[k], \label{eq:est_barzeta}
\end{align}
where $\bm{A}_{f}:= A \otimes I_{N-1}$, $\bm{B}_{f} := B \otimes I_{N-1}$, $\bm{C}_{f} := C \otimes I_{N-1}$, and $y_{\beta}[k]$ is the measurements corresponding to $\zeta_{\beta}[k]$, $\breve{\eta}_{\beta}[k] := (I_2 \otimes V_{\beta}) \eta[k]$.
The observer gains are
\begin{align*}
\bm{H}_{f} &:= \mathcal{F}(\bm{A}_{f},\bm{C}_{f},(I_2 \otimes V_{\beta})\bm{Q}(\tau)(I_2 \otimes V_{\beta}^{\sf T}), R \otimes I_{N-1}), \\
\bm{H}_{\bar{f}} &:= (I_2 \otimes q_{\tau}^{\sf T} V_{\beta T}^+)\bm{H}_{f}.
\end{align*}
Using the estimates of \eqref{eq:est_barzeta}, we apply an intermittent control strategy
\begin{equation}\label{eq:control_reg}
\bar{\eta}[k]:= \simode{
- F_{f} &\hat{\bar{z}}^-_{T}[k], ~&& k= K \ell, ~K \in \mathbb Z, \\
&0, ~&&\text{otherwise}.
}
\end{equation}
This control input is common to all MACs. 
Then, \textbf{Requirement \uppercase\expandafter{\romannumeral2}} in emergency operating mode is satisfied as follows. 
\begin{theorem}\label{theo:reg_long}
Consider the system \eqref{eq:ensmmld} under synchronization algorithm in Theorem \ref{theo:sync} with $D = \varGamma(\tau)$ and the control strategy \eqref{eq:control_reg}.
For a given $T >0$, suppose that the feedback gain $F_{f}$ is structured as 
\begin{equation}\label{eq:F_f}
    F_{f} = \gamma_{f} \mat{ \frac{\alpha_{f}}{T} ~&~ 1 },
\end{equation}
where $\gamma_{f}$ and $\alpha_{f}$ are scalar parameters.
Then 
\begin{equation}\label{eq:delta_r}
     \{x_i[k] -  r[k]\}_{k \in \ell\mathbb{Z}} \in \mathds{G}, \quad \forall i \in \mathcal{N},
\end{equation}
for the free-running dynamics $\Phi(q_{T})$ in \eqref{eq:syncdes} if and only if 
\begin{equation}\label{eq:con_Fr}
0< \gamma_{f} <\frac{4}{2+\alpha_{f}}, \quad
\alpha_{f} >0.
\end{equation}
\end{theorem}
\begin{proof}
We first prove that \eqref{eq:est_barzeta} gives a valid estimate of $(I_2 \otimes q_{\tau}^{\sf T} V_{\beta T}^+) \zeta_{\beta}[k]$.
Although the edge states $ \zeta_{\beta}[k]$ are distributed throughout the network, it has been shown in \eqref{eq:den_edge} that their measurements are decoupled. 
Therefore, a central estimator for $ \zeta_{\beta}[k]$ can be constructed as in \eqref{eq:est_zeta_beta}. Furthermore, since the floating control signal $\bar{\eta}[k]$ belongs to $\sfim(I_2 \otimes \mathds{1}_N^{\sf T})$, it does not affect $\zeta_{\beta}[k]$. 
Therefore, we can claim that the estimation error of \eqref{eq:est_zeta_beta} is guaranteed to converge.

On the other hand, according to \eqref{eq:def_geninv}, we know that 
$\sfim \mathds{1}_N = \sfker V,~\sfker q_{\tau} = \sfim V_{\beta \tau}^+,~\sfker q_{T} = \sfim V_{\beta T}^+$.
In the context of projection matrix, this implies that
\[
\mathds{1}_N q_{\tau}^T + V_{\beta \tau}^+ V_{\beta} = I_N, \quad \mathds{1}_N q_{T}^T + V_{\beta T}^+ V_{\beta} = I_N.
\]
Therefore, we have
\begin{equation}\label{eq:qV_relation}
q_{\tau}^{\sf T} V_{\beta T}^+ = q_{\tau}^{\sf T} (I_N - \mathds{1}_N q_{T}^T) V_{\beta \tau}^+  = - q_{T}^{\sf T} V_{\beta \tau}^+.
\end{equation}
Together with the fact that $\eta[k] \in \sfker(I_2 \otimes q_{\tau}^{\sf T})$, we get
\[
(I_2 \otimes q_{\tau}^{\sf T} V_{\beta T}^+)~ \breve{\eta}_{\beta}[k] = 0, \quad \forall k \in \mathbb{Z}.
\]
So the synchronization control-related term does not appear in \eqref{eq:est_barzeta}. This claims the conclusion.

With such a valid estimator, it is not hard to see that by choosing an appropriate gain that satisfies condition \eqref{eq:con_Fr},  $\bar{z}^-_{T}[k]$ is stabilized, similar to what was proved in Theorem \ref{theo:anchor}. 
Thus, the synchronized state follows $\bar{z}_{T}[k]$. 

Furthermore, we examine the dynamics of $\bar{z}_{T}[k]$ that
\[
\bar{z}_{T}[k+1] = A \bar{z}_{T}[k] + B \bar{\eta}_{T}[k] + (I_2 \otimes q_{T}^{\sf T}) v[k],
\]
where $\bar{\eta}_{T}[k]$ denotes its equivalent control input. 
According to the expansion equation \eqref{eq:control_expan} and similar analysis as \eqref{eq:basisqinfq}, we know
\[
\bar{\eta}_{T}[k] = (I_2 \otimes q_{T}^{\sf T} V_{\beta \tau}^+) \check{\eta}[k] + \bar{\eta}[k].
\]
The first term is equal to $-(I_2 \otimes q_{\tau}^{\sf T} V_{\beta {T}}^+) \check{\eta}[k]$ by \eqref{eq:qV_relation}, while the second term is proportional to $ (I_2 \otimes q_{\tau}^{\sf T} V_{\beta T}^+) \zeta_{\beta}[k]$ as proved earlier.
Recall that $\check{\eta}[k]$ is the control input used to stabilize $\zeta_{\beta}[k]$, and both $\zeta_{\beta}[k]$ and $\check{\eta}[k]$ converge as proved in Theorem \ref{theo:sync}. 
Therefore, we conclude that $\{\bar{\eta}_{T}[k]\}_{k\in\mathbb{Z}}\in \mathds{G}$. 
In other words, the difference between $\bar{z}_{T}[k]$ and $r[k]$ of $\Phi(q_{T})$ converges. 
This completes the proof.
\hfill
 $\square$
\end{proof}

The intermittent floating control periodically maximizes the long-term frequency stability of the GTS, while maintaining its inherent short-term frequency stability. 
Consequently, the combination of Corollary \ref{theo:reg_short} and Theorem \ref{theo:reg_long} 
gives an optimal GTS over both short and long periods.

\begin{figure}[t]
\centering
\includegraphics[scale=0.4]{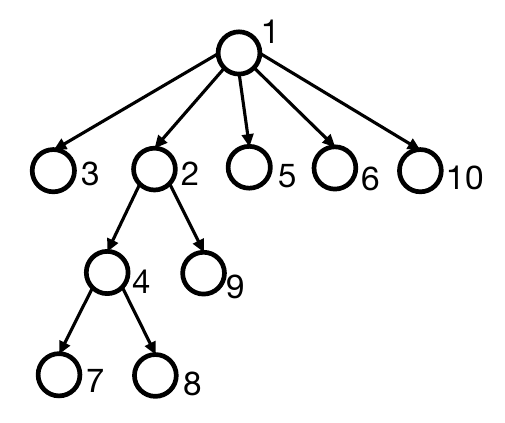}
\caption{A directed spanning tree of the network topology.}
\label{fig:stree}
\end{figure}

\textbf{Illustrative Example:}
We demonstrate the optimal floating control of the GTS in the emergency operating mode.
The simulation setup is the same as in Section~\ref{sec:Anchoring}.
To construct the state estimator in \eqref{eq:est_zeta_beta}, we consider a directed spanning tree of the network topology, as shown in Fig.~\ref{fig:stree}, where the node labels are inherited from Fig.~\ref{fig:topology}.

\begin{figure}[t]
\centering
\includegraphics[scale=0.38]{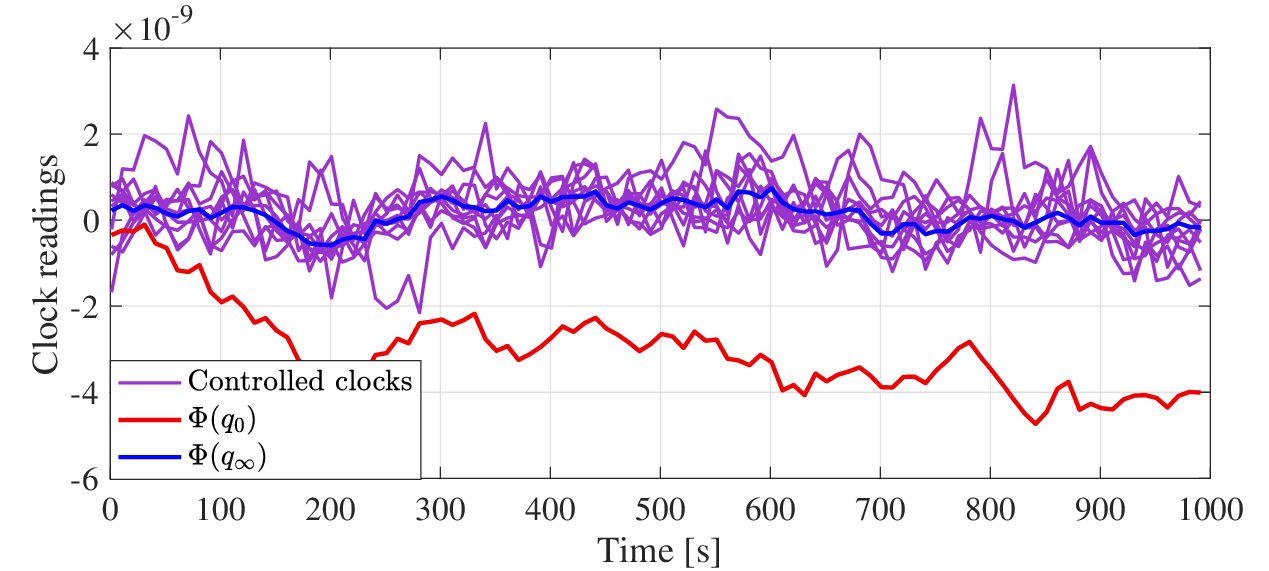}
\vspace{1em}
\includegraphics[scale=0.38]{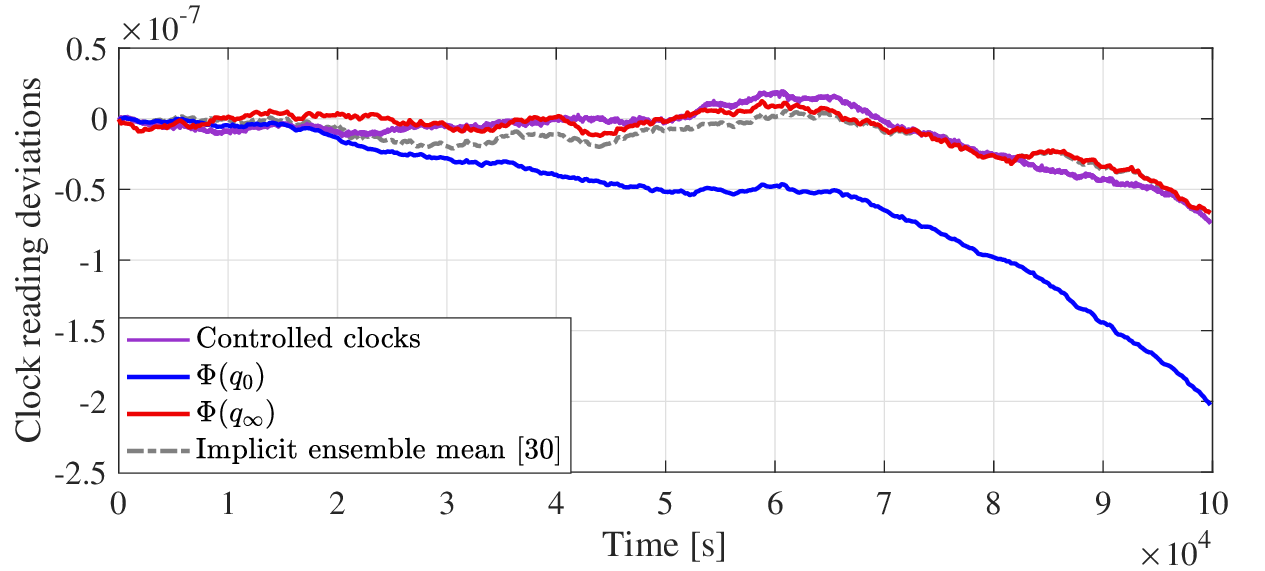}
\caption{Optimal floating MACs in emergency operating mode.}
\label{fig:State_1e3_5}
\end{figure}

The clock reading deviations for the time period $k \in [1,10^3]$ and $k \in [1,10^5]$ are shown in Fig.~\ref{fig:State_1e3_5}. 
In both subfigures, the purple lines represent the MACs controlled by both distributed synchronization and optimal floating strategies. 
The blue lines and red lines correspond to the virtual trajectories generated by $\Phi(q_{0})$ and $\Phi(q_{\infty})$, respectively, where $q_{0}$ and $q_{\infty}$ are given in \eqref{eq:q_zero} and \eqref{eq:q_inf}. 
With respect to the short period in the upper subfigure, the MACs are synchronized but floating, i.e., they follow $\Phi(q_{0})$. 
While for the longer time period in the lower subfigure, the MACs are explicitly regulated to  $\Phi(q_{\infty})$ with the maximum long-term frequency stability instead. 
Obviously, this reduces the rate of divergence, resulting in more desirable timing in the event of GNSS failures.
Note that this optimal timing in both the short and long term is realized here even without the standard time reference.

\blue{
The centralized time scale generation method, the implicit ensemble mean (IEM) \cite{galleani2010time}, 
is used as a baseline to evaluate the effectiveness of the proposed distributed architecture. 
It predicts the MAC readings via a standard Kalman filter based on inter-clock measurements. 
As indicated by \cite{ishizaki2026explicit}, this method inherently optimizes long-term ensemble performance. 
This behavior is evident in the lower subfigure, where the IEM time scale eventually aligns with $\Phi(q_{\infty})$. 

To provide a quantitative comparison, two statistical indicators are computed: the mean deviation (MD) and the root-mean-square error (RMSE). 
For a clock reading deviation sequence $\{h_{i}[0],\ldots,h_{i}[M]\}$, these metrics are defined as
\[
\mathrm{MD} = \frac{1}{M}\sum_{k=1}^{M} h_{i}[k],
\quad
\mathrm{RMSE} = 
\sqrt{\frac{1}{M}\sum_{k=1}^{M} (h_{i}[k])^2 }.
\]
The resulting values of the IEM baseline and MAC $1$ under the proposed method (prop) are
\begin{align*}
&\mathrm{MD}_{\mathrm{IEM}} = -6.801 \times 10^{-8},&\mathrm{RMSE}_{\mathrm{IEM}} = 1.318 \times 10^{-7}, \\
&\mathrm{MD}_{\mathrm{prop}} = -6.845 \times 10^{-8},&\mathrm{RMSE}_{\mathrm{prop}} = 1.410 \times 10^{-7}. 
\end{align*}
These results demonstrate that the combination of distributed synchronization control and optimal floating control yields performance comparable to that of the centralized IEM method.
}

\begin{figure}[t]
\centering
\includegraphics[scale=0.4]{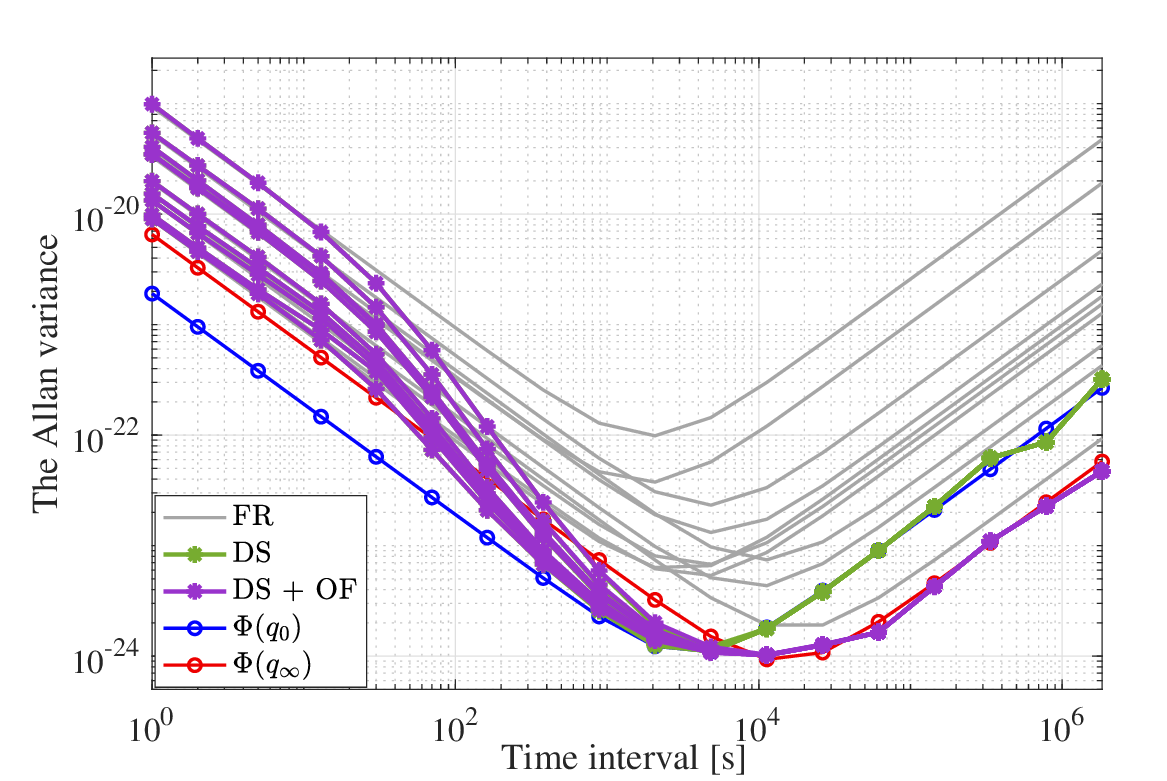}
\caption{The AVAR of the MACs that are free-running (FR), controlled by the distributed synchronization strategy alone (DS), controlled by both the distributed synchronization and the optimal floating control strategies (DS + OF), and the free-running ensemble mean $\Phi(q_{0})$ and $\Phi(q_{\infty})$.}
\label{fig:float_avar}
\end{figure}

The AVAR curves of the MACs under both distributed synchronization control and optimal floating control are plotted with the purple lines in Fig.~\ref{fig:float_avar}. 
We can see that the long-term frequency stability is  maximized while maintaining the optimal short-term frequency stability.

\blue{
We further examine the mode-switching scenario in which GNSS failure occurs. 
For $k \in [0,1000)$, GNSS signals are available and the supervisor applies anchoring control.
At $k=1000$, the GNSS signals are suddenly lost, after which the supervisor switches from the normal operating mode to the emergency mode.  
As shown in Fig. \ref{fig:State_switch}, the MACs remain synchronized throughout the transition, and the switch does not introduce any noticeable disturbance. 
The GTS preserves accuracy before the failure and then floats optimally thereafter as expected.
These results fully demonstrate the effectiveness and robustness of the proposed hierarchical distributed timing architecture.
\begin{figure}[t]
\centering
\includegraphics[scale=0.4]{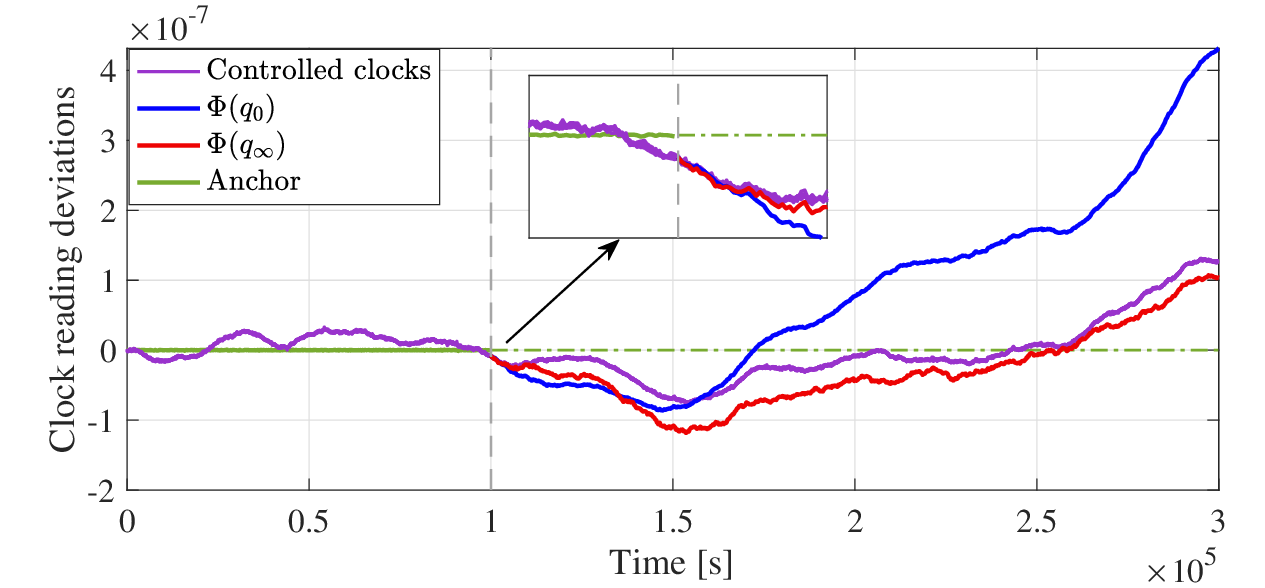}
\caption{Switching between normal and emergency operating modes. GNSS failure at $k=1000$.}
\label{fig:State_switch}
\end{figure}
}

\section{Conclusion}
\label{sec:Conclusion}
In this paper, we have presented a hierarchical distributed architecture for achieving stable and accurate atomic timing. 
Using a state-space expansion method, the desired timing objectives are naturally decoupled into distributed synchronization and anchoring/floating control tasks. 
The proposed architecture demonstrates that a well-coordinated combination of distributed control and supervisory control can maintain long-term, high-precision timing in the presence of GNSS uncertainties.

The primary focus of this work is to establish the fundamental framework of distributed atomic timing. 
\blue{
Future research directions include the development of systematic strategies for handling the gradual degradation of GNSS signals, as well as the analysis and accommodation of network topology variations arising from link disruptions or partial MAC failures. 
}
Additional topics of interest include optimizing control performance with respect to energy efficiency and extending the proposed framework to multiple interacting atomic clock ensembles.

\begin{ack} 
This work is supported by research and development conducted by the Ministry of Internal Affairs and Communications (MIC) under its ``Research and Development for Expansion of Radio Resources (JPJ000254)'' program.
\end{ack}

\bibliographystyle{unsrt}
\bibliography{ref}

@book{bregni2002synchronization,
	author = {Bregni, Stefano},
	publisher = {John Wiley \& Sons, Inc.},
	title = {Synchronization of digital telecommunications networks},
	year = {2002}}

@article{zhang2013time,
	author = {Zhang, Zhenghao and Gong, Shuping and Dimitrovski, Aleksandar D and Li, Husheng},
	journal = {IEEE Transactions on Smart Grid},
	number = {1},
	pages = {87--98},
	publisher = {IEEE},
	title = {Time synchronization attack in smart grid: {I}mpact and analysis},
	volume = {4},
	year = {2013}}

@article{yiugitler2020overview,
	author = {Yi{\u{g}}itler, H{\"u}seyin and Badihi, Behnam and J{\"a}ntti, Riku},
	journal = {Sensors},
	number = {20},
	pages = {5928},
	publisher = {MDPI},
	title = {Overview of time synchronization for {IoT} deployments: {C}lock discipline algorithms and protocols},
	volume = {20},
	year = {2020}}

@article{hasan2018time,
	author = {Hasan, Khondokar Fida and Wang, Charles and Feng, Yanming and Tian, Yu-Chu},
	journal = {Vehicular Communications},
	pages = {39--51},
	publisher = {Elsevier},
	title = {Time synchronization in vehicular ad-hoc networks: {A} survey on theory and practice},
	volume = {14},
	year = {2018}}

@inproceedings{montenbruck2020comparing,
	author = {Montenbruck, Oliver and Steigenberger, Peter and Hauschild, Andr{\'e}},
	booktitle = {2020 IEEE/ION Position, Location and Navigation Symposium (PLANS)},
	organization = {IEEE},
	pages = {407--418},
	title = {Comparing the `{B}ig 4'-{A} User's View on {GNSS} Performance},
	year = {2020}}

@article{jaduszliwer2021past,
	author = {Jaduszliwer, Bernardo and Camparo, James},
	journal = {GPS Solutions},
	pages = {1--13},
	publisher = {Springer},
	title = {Past, present and future of atomic clocks for {GNSS}},
	volume = {25},
	year = {2021}}

@article{allan1966statistics,
	author = {Allan, David W},
	journal = {Proceedings of the IEEE},
	number = {2},
	pages = {221--230},
	publisher = {IEEE},
	title = {Statistics of atomic frequency standards},
	volume = {54},
	year = {1966}}

@article{ishizaki2023higher,
	author = {Ishizaki, Takayuki and Ichimura, Taichi and Kawaguchi, Takahiro and Yano, Yuichiro and Hanado, Yuko},
	journal = {Metrologia},
	number = {1},
	pages = {015003},
	publisher = {IOP Publishing},
	title = {Higher-order {A}llan variance for atomic clocks of arbitrary order: mathematical foundation},
	volume = {61},
	year = {2023}}

@article{kitching2018chip,
	author = {Kitching, John},
	journal = {Applied Physics Reviews},
	number = {3},
	publisher = {AIP Publishing},
	title = {Chip-scale atomic devices},
	volume = {5},
	year = {2018}}

@article{gill2005optical,
	author = {Gill, Patrick},
	journal = {Metrologia},
	number = {3},
	pages = {S125--S137},
	title = {Optical frequency standards},
	volume = {42},
	year = {2005}}

@article{bandi2024comprehensive,
	author = {Bandi, Thejesh N},
	journal = {Demo Journal},
	number = {1},
	pages = {40--50},
	title = {A comprehensive overview of atomic clocks and their applications},
	volume = {1},
	year = {2024}}

@article{ishizaki2026explicit,
  title={Explicit ensemble mean clock synchronization for optimal atomic time scale generation},
  author={Ishizaki, Takayuki and Kawaguchi, Takahiro and Yano, Yuichiro and Hanado, Yuko},
  journal={Metrologia},
  volume={63},
  number={2},
  pages={025010},
  year={2026},
  publisher={IOP Publishing}}

@book{banerjee2023introduction,
	author = {Banerjee, Parameswar and Matsakis, Demetrios},
	publisher = {Springer},
	title = {An introduction to modern timekeeping and time transfer},
	year = {2023}}

@article{winzer2018fiber,
	author = {Winzer, Peter J and Neilson, David T and Chraplyvy, Andrew R},
	journal = {Optics Express},
	number = {18},
	pages = {24190--24239},
	publisher = {OSA},
	title = {Fiber-optic transmission and networking: the previous 20 and the next 20 years},
	volume = {26},
	year = {2018}}

@article{coleman2020autonomous,
	author = {Coleman, Michael J and Beard, Ronald L},
	journal = {NAVIGATION: Journal of the Institute of Navigation},
	number = {2},
	pages = {333--346},
	publisher = {Institute of Navigation},
	title = {Autonomous clock ensemble algorithm for {GNSS} applications},
	volume = {67},
	year = {2020}}

@book{brown1991theory,
	author = {Brown, Kenneth R},
	publisher = {Institute of Navigation},
	title = {The theory of the GPS composite clock},
	year = {1991}}

@article{kunzi2023precise,
	author = {Kunzi, Florian and Montenbruck, Oliver},
	journal = {GPS Solutions},
	number = {4},
	pages = {165},
	publisher = {Springer},
	title = {Precise disciplining of a chip-scale atomic clock using {PPP} with broadcast ephemerides},
	volume = {27},
	year = {2023}}

@article{zidan2020gnss,
	author = {Zidan, Jasmine and Adegoke, Elijah I and Kampert, Erik and Birrell, Stewart A and Ford, Col R and Higgins, Matthew D},
	journal = {IEEE Access},
	pages = {153960--153976},
	publisher = {IEEE},
	title = {{GNSS} vulnerabilities and existing solutions: {A} review of the literature},
	volume = {9},
	year = {2020}}

@article{galleani2003use,
	author = {Galleani, Lorenzo and Tavella, Patrizia},
	journal = {Metrologia},
	number = {3},
	pages = {S326},
	publisher = {IOP Publishing},
	title = {On the use of the {K}alman filter in timescales},
	volume = {40},
	year = {2003}}

@article{galleani2010time,
	author = {Galleani, Lorenzo and Tavella, Patrizia},
	journal = {IEEE Control Systems Magazine},
	number = {2},
	pages = {44--65},
	publisher = {IEEE},
	title = {Time and the {K}alman filter},
	volume = {30},
	year = {2010}}

@article{liu2020secure,
	author = {Liu, Yin-Chen and Bianchin, Gianluca and Pasqualetti, Fabio},
	journal = {Automatica},
	pages = {108655},
	publisher = {Elsevier},
	title = {Secure trajectory planning against undetectable spoofing attacks},
	volume = {112},
	year = {2020}}

@article{radovs2024recent,
	author = {Rado{\v{s}}, Katarina and Brki{\'c}, Marta and Begu{\v{s}}i{\'c}, Dinko},
	journal = {Sensors},
	number = {13},
	pages = {4210},
	publisher = {MDPI},
	title = {Recent advances on jamming and spoofing detection in {GNSS}},
	volume = {24},
	year = {2024}}

@inproceedings{cash2018microsemi,
	author = {Cash, Peter and Krzewick, Will and Machado, Paul and Overstreet, K Richard and Silveira, Mike and Stanczyk, Matt and Taylor, Dwayne and Zhang, Xianli},
	booktitle = {2018 European Frequency and Time Forum (EFTF)},
	organization = {IEEE},
	pages = {65--71},
	title = {Microsemi chip scale atomic clock ({CSAC}) technical status, applications, and future plans},
	year = {2018}}

@article{he2017accurate,
	author = {He, Jianping and Duan, Xiaoming and Cheng, Peng and Shi, Ling and Cai, Lin},
	journal = {Automatica},
	pages = {350--358},
	publisher = {Elsevier},
	title = {Accurate clock synchronization in wireless sensor networks with bounded noise},
	volume = {81},
	year = {2017}}

@article{kadowaki2014event,
	author = {Kadowaki, Yuki and Ishii, Hideaki},
	journal = {IEEE Transactions on Automatic Control},
	number = {8},
	pages = {2266--2271},
	publisher = {IEEE},
	title = {Event-based distributed clock synchronization for wireless sensor networks},
	volume = {60},
	year = {2014}}

@article{wang2022consensus,
	author = {Wang, Heng and Gong, Pengfei and Li, Min},
	journal = {Automatica},
	pages = {110045},
	publisher = {Elsevier},
	title = {Consensus-based time synchronization via sequential least squares for strongly rooted wireless sensor networks with random delays},
	volume = {136},
	year = {2022}}

@article{chen2019bit,
	author = {Chen, Jiayu and Ling, Qiang},
	journal = {IEEE Transactions on Cybernetics},
	number = {2},
	pages = {984--993},
	publisher = {IEEE},
	title = {Bit-rate conditions for the consensus of quantized multiagent systems with network-induced delays based on event triggering},
	volume = {51},
	year = {2019}}

@article{galleani2008tutorial,
	author = {Galleani, Lorenzo},
	journal = {Metrologia},
	number = {6},
	pages = {S175},
	publisher = {IOP Publishing},
	title = {A tutorial on the two-state model of the atomic clock noise},
	volume = {45},
	year = {2008}}

@article{olfati2004consensus,
	author = {Olfati-Saber, Reza and Murray, Richard M},
	journal = {IEEE Transactions on Automatic Control},
	number = {9},
	pages = {1520--1533},
	publisher = {IEEE},
	title = {Consensus problems in networks of agents with switching topology and time-delays},
	volume = {49},
	year = {2004}}

@inproceedings{dey2024clock,
	author = {Dey, Priyanka and Murugesan, Sathishkumar and Kawaguchi, Takahiro and Yano, Yuichiro and Hanado, Yuko and Ishizaki, Takayuki},
	booktitle = {2024 European Control Conference (ECC)},
	organization = {IEEE},
	pages = {2132--2137},
	title = {Clock Steering Techniques for Atomic Clocks of Arbitrary Order},
	year = {2024}}

@book{anderson2005optimal,
	author = {Anderson, Brian DO and Moore, John B},
	publisher = {Courier Corporation},
	title = {Optimal filtering},
	year = {2005}}

@article{mills1991internet,
	author = {Mills, David L},
	journal = {IEEE Transactions on Communications},
	number = {10},
	pages = {1482--1493},
	publisher = {Ieee},
	title = {Internet time synchronization: the network time protocol},
	volume = {39},
	year = {1991}}

@inproceedings{serrano2013white,
	author = {Serrano, J and Cattin, M and Gousiou, E and van der Bij, E and W{\l}ostowski, Tomasz and Daniluk, G and Lipi{\'n}ski, Maciej Marek},
	booktitle = {International Beam Instrumentation Conference},
	organization = {Diamond Light Source},
	title = {The White Rabbit Project},
	year = {2013}}

@article{pi2017effects,
	author = {Pi, Xiaoqing and Iijima, Byron A and Lu, Wenwen},
	journal = {Navigation: Journal of the Institute of Navigation},
	number = {1},
	pages = {3--22},
	publisher = {Wiley Online Library},
	title = {Effects of ionospheric scintillation on {GNSS}-based positioning},
	volume = {64},
	year = {2017}}

@article{koppang2016state,
	author = {Koppang, Paul A},
	journal = {Metrologia},
	number = {3},
	pages = {R60},
	publisher = {IOP Publishing},
	title = {State space control of frequency standards},
	volume = {53},
	year = {2016}}

@article{marlow2021review,
	author = {Marlow, Bonnie L Schmittberger and Scherer, David R},
	journal = {IEEE Transactions on Ultrasonics, Ferroelectrics, and Frequency Control},
	number = {6},
	pages = {2007--2022},
	publisher = {IEEE},
	title = {A review of commercial and emerging atomic frequency standards},
	volume = {68},
	year = {2021}}

\end{document}